\documentclass[useAMS,usenatbib]{mn2e}
\usepackage{graphicx,amsmath,multirow,amssymb}
\usepackage{subfig}
\usepackage{natbib}
\newcommand{\comment}[1]{}

\def\simgt{\lower.5ex\hbox{$\; \buildrel > \over \F sim \;$}}
\def\simlt{\lower.5ex\hbox{$\; \buildrel < \over \sim \;$}}

\title[M-stars in the LMC]{Characterization of M-stars in the LMC in the JWST era.}

\author[E. Marini et al.]{
E. Marini,$^{1,2}$\thanks{E-mail: ester.marini@uniroma3.it}
F. Dell'Agli,$^{2}$,
M. Di Criscienzo$^{2}$, D. A. Garc\'{\i}a--Hern\'andez$^{3,4}$, 
\newauthor
P. Ventura$^{2}$, M. A. T. Groenewegen$^{5}$,  L. Mattsson$^{6}$, D. Kamath$^{7}$, 
S. Puccetti$^{8}$, \newauthor
M. Tailo$^{9}$, and E. Villaver$^{10}$
\\
$^{1}$Dipartimento di Matematica e Fisica, Universit\'a degli Studi Roma Tre, via
della Vasca Navale 84, 00100, Roma\\
$^{2}$INAF, Osservatorio Astronomico di Roma, Via Frascati 33, 00077, Monte Porzio Catone, Italy\\
$^{3}$Instituto de Astrof\'{\i}sica de Canarias (IAC), E-38200 La Laguna, Tenerife, Spain \\
$^{4}$Departamento de Astrof\'{\i}sica, Universidad de La Laguna (ULL), E-38206 La Laguna, Tenerife, Spain \\
$^{5}$Koninklijke Sterrenwacht van Belgi{\"e}, Ringlaan 3, 1180 Brussels, Belgium\\
$^{6}$Nordita, KTH Royal Institute of Technology and Stockholm University, Roslagstullsbacken 23, SE-106 91 Stockholm, Sweden\\
$^{7}$Department of Physics and Astronomy, Macquarie University, Sydney, NSW 2109, Australia\\
$^{8}$ASI, Via del Politecnico, 00133 Roma, Italy \\
$^{9}$Dipartimento di Fisica e Astronomia 'Galileo Galilei', Univ. di Padova, Vicolo dell?Osservatorio 3, I-35122 Padova, Italy\\
$^{10}$Departamento de Fisica Teorica, Universidad Autonoma de Madrid, Cantoblanco 28049 Madrid, Spain\\
}

\begin{document}

\date{Accepted, Received; in original form }

\pagerange{\pageref{firstpage}--\pageref{lastpage}} \pubyear{2012}

\maketitle

\label{firstpage}

\begin{abstract}
We study the M-type asymptotic giant branch (AGB) population of the Large Magellanic Cloud
(LMC) by characterizing the individual sources in terms of the main properties of the 
progenitors and of the dust present in the circumstellar envelope. To this aim
we compare the combination of the spectroscopic and photometric data collected by 
{\itshape Spitzer}, complemented by additional photometric results available in the literature, with 
results from AGB modelling that include the description of dust formation in the wind.
To allow the interpretation of a paucity stars likely evolving through the post-AGB phase,
we extended the available evolutionary sequences to reach the PN phase. The main motivation
of the present analysis is to prepare the future observations of the evolved stellar
populations of Local Group galaxies that will be done by the James Webb Space Telescope
({\itshape JWST}), by identifying the combination of filters that will maximize the possibilities of 
characterizing the observed sources. The present results show that for the M-star case the
best planes to be used for this purpose are the colour magnitude ([F770W]-[F2550W], [F770W]) 
and (K$_S$-[F770W], [F770W]) planes. In these observational diagrams the sequences of 
low-mass stars evolving in the AGB phases before the achievement of the C-star stage and
of massive AGBs experiencing hot bottom burning are clearly separated and peculiar sources,
such as post-AGB, dual-dust chemistry and iron-dust stars can be easily identified.
\end{abstract}




\section{Introduction}
The stars evolving through the AGB provide an important
feedback on their host system. During this phase they lose their entire external mantle,
ejecting into the interstellar medium (ISM) large quantities of gas, partly contaminated
by internal nucleosynthesis and mixing processes. Their circumstellar envelope is a 
favourable environment to the formation of dust \citep{woitke99}, which is lost into their 
surroundings, owing to the effects of the pulsation and of the radiation pressure acting 
on the dust grains. In this way, AGB stars participate to the dust cycle of the host 
galaxy \citep{javadi16, li19}. 

Addressing these important topics is now possible thanks to the latest generation of 
models, that couple the simulation of the evolution of the central star and the 
description of the dust formation process, which, in turn, is coupled self-consistently 
with the dynamics of the wind \citep{fg02, fg06}. 
These preliminary investigations have proven extremely useful to foresee the amount 
of dust produced by stars of different mass and chemical composition, 
across the various stages of the AGB evolution \citep{flavia17, marcella13, nanni13, 
nanni14, ventura12, ventura14, ventura18}. 

Despite these important progresses, the estimates of dust yields by AGB stars are still
affected by several uncertainties, which can be broadly grouped into 3 distinct categories:
a) the AGB evolution is strongly determined by the efficiency of two physical phenomena,
still poorly known from first principles, namely convection and mass loss \citep{ventura05a,
ventura05b, karakas14}; b) the dynamics of the wind is characterized by the formation of 
shocks, which provoke significant deviations with respect to the isotropic, 
stationary schematization currently used \citep{bowen88, Cherchneff}; c) Dust 
production mechanism itself is still affected by several uncertainties, related to the 
scarce knowledge of the sticking coefficients of the molecules on the solid particles and 
to the formation entalpies of some solid compounds \citep{fg06, gail13}.

On the observational ground, photometry and spectroscopy in the infrared (IR) domain prove 
valuable tools to improve the
understanding of the evolution and of the dust production mechanism by AGB stars. Indeed, 
when dust is formed a significant fraction of the overall energy released is emitted in
the IR spectral region. Furthermore, the IR spectrum is characterized by various
features, each associated to a specific dust species; this is important to deduce the
mineralogy of the dust formed.

The LMC has been so far the best laboratory to test 
AGB evolution theories. This is due to its relative proximity 
\citep[$\sim 50$ Kpc,][]{feast99} and low average reddening 
\citep[$E(B-V) \sim 0.075$,][]{schlegel98}, which allowed the observation of
the evolved stellar population by means of several surveys. The most recent and complete 
exploration has been achieved via the Surveying the Agents of a Galaxy's Evolution 
Survey (SAGE), with the Spitzer Space Telescope \citep{meixner06}, that provided IR data
taken with the InfraRed Array Camera (IRAC, with filters centered at $3.6$, $4.5$, $5.8$ 
and $8.0~\mu$m) and the Multi-band Imaging Photometer (MIPS, with filter centered
at $24~\mu$m) of $\sim 6.5$ million sources, $\sim 17000$ out of which were classified as AGB
stars by \citet{riebel10}.

The availability of this robust body of observational data has allowed the study of the dust 
enrichment from stellar sources from two different perspectives. Several authors used 
synthetic spectra, obtained by varying the parameters of the central object and the dust 
composition, to reproduce the position of the observed sources in the observational planes 
built with the IRAC and MIPS filters \citep{srinivasan09, srinivasan10, srinivasan11, riebel12}. 
A different and complementary approach was followed by \citet{flavia14a, flavia15a}, 
who used stellar evolutionary tracks to characterize the individual sources, in terms 
of mass, chemical composition and formation epoch of the progenitors, and of the amount 
and mineralogy of the dust in the circumstellar envelope. \citet{ambra19} used a
similar analysis to derive an estimate of the overall dust production rate by evolved stars
in the Magellanic Clouds (hereinafter MC).

A further step towards the interpretation of the IR observations of LMC stars is possible
via the analysis of spectroscopic data taken with the {\itshape Spitzer}'s Infrared Spectrograph (IRS), which
provided detailed mid-IR spectral distribution of more than 1000 point sources
in the LMC. The fit of the IR spectra allows a wider and deeper exploration of the various
factors affecting the spectral energy distribution, in comparison to the analysis based on 
the different magnitudes. This approach was followed by \citet{jones14} and 
Groenewegen \& Sloan (2018, GS18) to characterize
oxygen-rich AGB stars.

In the near future, the studies aimed at understanding how dust production in the 
envelope of AGB stars works will receive a robust push, with the launch of the {\itshape JWST}, 
that will revolutionize our understanding of the evolved stellar 
populations in the local Universe. The large aperture (6.5 m) and the subarsecond spatial 
resolution will allow the study of resolved dusty stellar populations at moderate and large
distances, up to $\sim 4$ Mpc \citep{jones17}. The Mid-Infrared Instrument (MIRI; Rieke et al. 2015), 
mounted onboard the {\itshape JWST}, will provide spectroscopy in the $5-28.5~\mu$m range 
\citep{bouchet15}, thus providing a unique opportunity to study the evolution of AGB 
stars and the dust formation process in their expanding wind, in a large variety of environments. 
Combination of near-IR and mid-IR data to study the evolved stellar population 
has been so far successfully applied to the MC. By analyzing 
the results from the DUSTiNGS survey, \citet{flavia16, flavia18b, flavia19} attempted a 
similar approach to study AGB stars in the Local Group galaxies IC1613, IC10 and Sextans A; 
however, these studies,
based on photometry results limited to wavelengths below $\sim 5~\mu$m, could not benefit
of the information of the long-wavelength domain of the mid-IR flux. The {\itshape JWST} data will
allow the application of this methodology to all the galaxies of the Local Group and
possibly beyond.

Against this background, we have started a research project with the goal of providing
a thorough interpretation of the IR data of the stars in the LMC that exhibit an IR excess.
In this paper we restrict our attention to the sample of oxygen-rich stars, while we will 
address carbon stars in a forthcoming work. In the wake of the analysis done in 
\citet{jones14}, we will combine results from IRAC and MIPS photometry with IRS data, 
in the attempt of giving an exhaustive characterization of the individual sources, in 
terms of the main properties of their progenitors, of the specific AGB stage they are 
evolving through and of the properties of the dust in their surroundings. 
To consider the whole spectral distribution, we will also take into account optical and
near IR  photometric data, when available.
This study represents 
a step forward with respect to the works by \citet{flavia14a, flavia15a}, which were based 
on photometric data only. The method followed here is different and complementary from 
\citet{jones14}, as it is based on results from AGB evolution and dust formation modelling.

Because the present work is projected into a {\itshape JWST} perspective, similarly to 
\citet{jones17} we will consider observational planes obtained by the combination of 
MIRI filters. Our goal is to select the observational planes that must be used in order
to obtain the most exhaustive characterization of the individual sources observed, in terms
of chemical composition, mass and formation epoch of the progenitors, of the degree of 
obscuration of the stars and of the mineralogy of the dust present in the circumstellar envelope.
A detailed care will be devoted to understand the planes where the different classes of the
sources observed can be easily distinguished and where the obscuration trends for 
oxygen-rich stars are most clearly defined. This step is crucial to set up a methodology
that will be extended to the galaxies in the Local Group, once the {\itshape JWST} data will become
available.

\section{AGB and dust formation modelling}
\label{models}

\begin{figure*}
\begin{minipage}{0.33\textwidth}
\resizebox{1.\hsize}{!}{\includegraphics{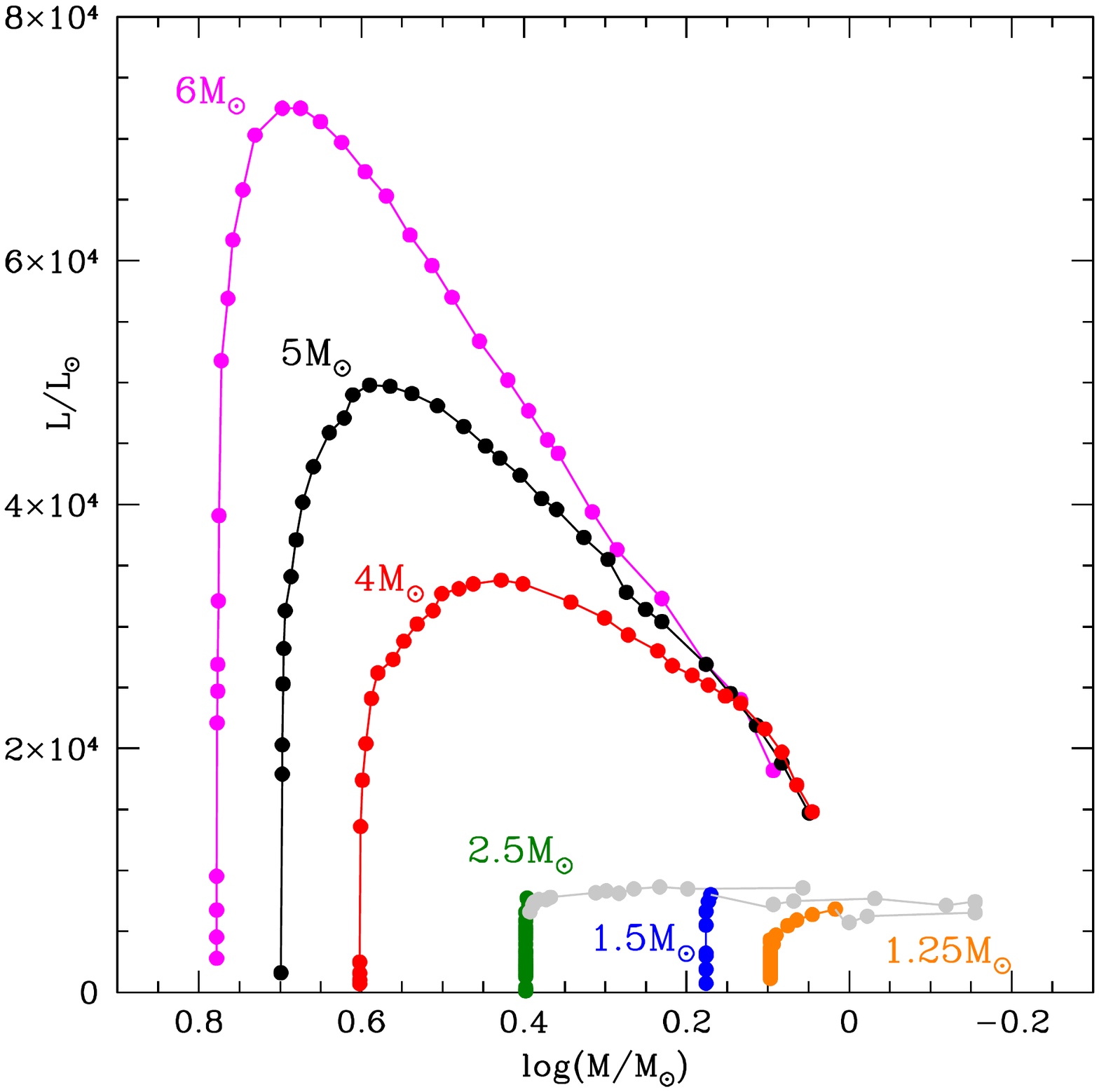}}
\end{minipage}
\begin{minipage}{0.33\textwidth}
\resizebox{1.\hsize}{!}{\includegraphics{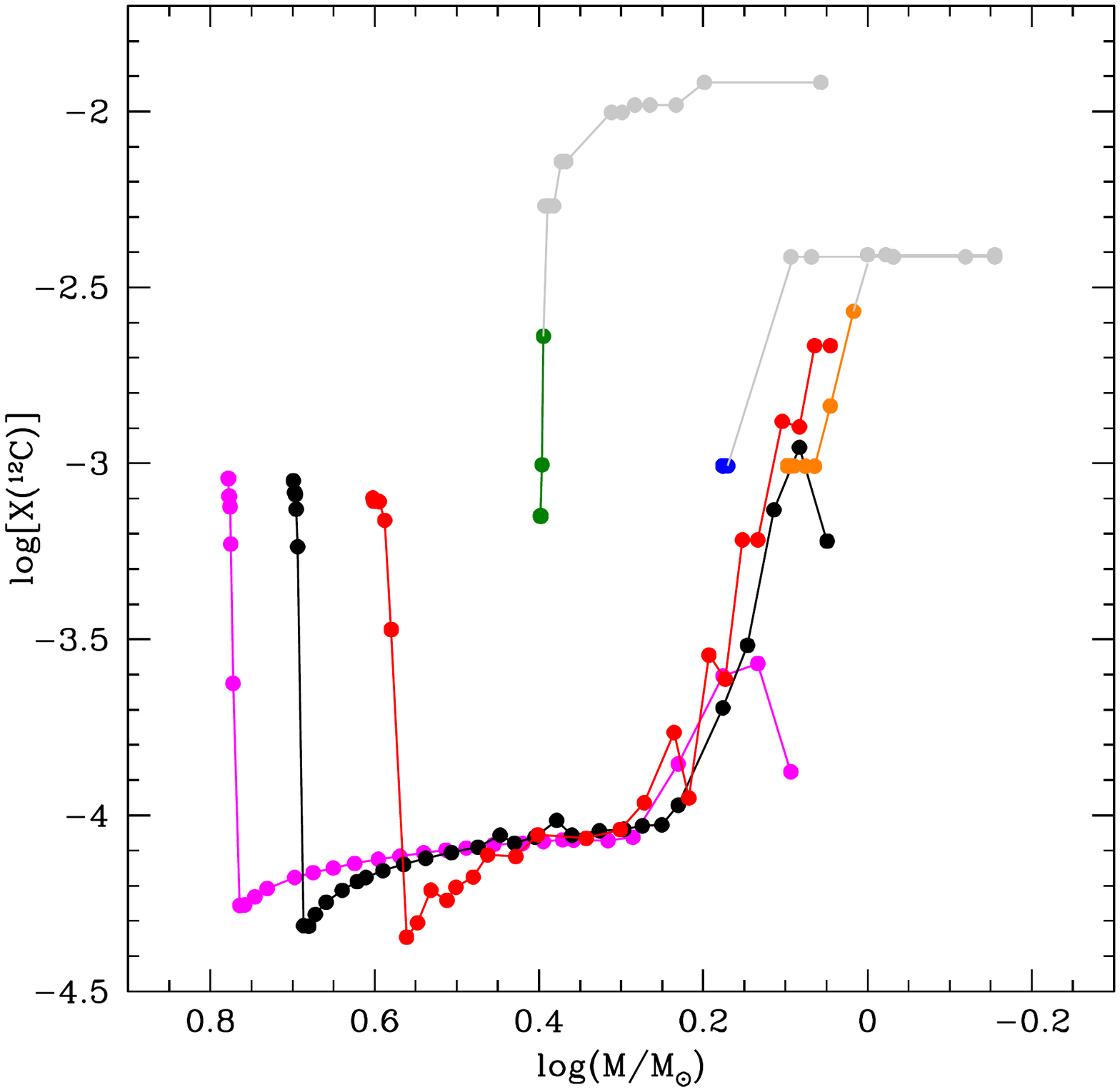}}
\end{minipage}
\begin{minipage}{0.33\textwidth}
\resizebox{1.\hsize}{!}{\includegraphics{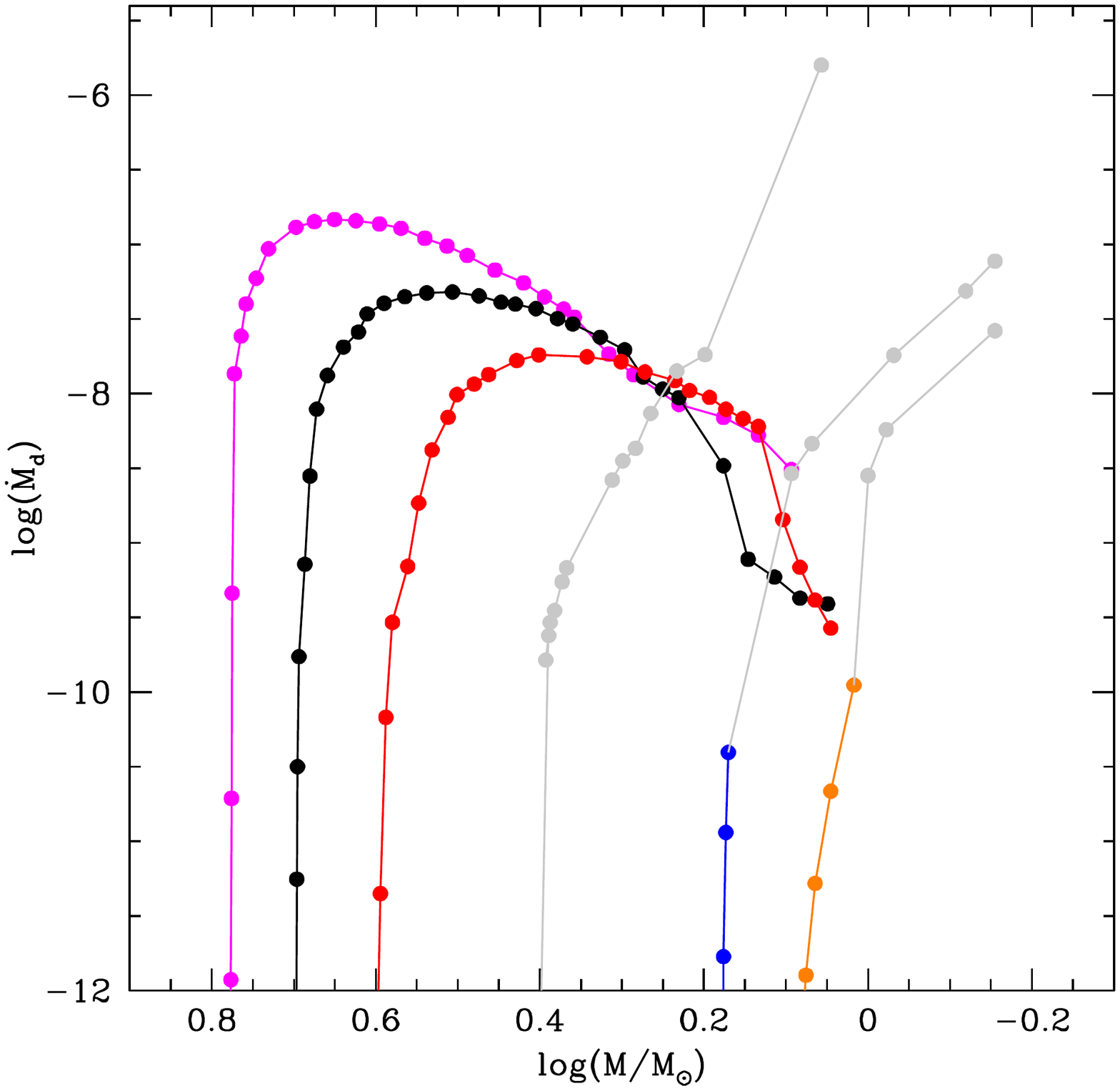}}
\end{minipage}
\vskip-40pt
\caption{The variation during the AGB phase of the luminosity (left panel), surface
carbon mass fraction (middle) and dust production rate (right),
as a function of the current mass of the star
(shown in a logarithmic scale), of stars of different mass and metallicity $Z=8\times 10^{-3}$. 
The color-coding is as follows: orange points - $1.25~M_{\odot}$; blue - $1.5~M_{\odot}$;
green - $2.5~M_{\odot}$; red - $4~M_{\odot}$; black - $5~M_{\odot}$; 
magenta - $6~M_{\odot}$; The points shown in grey refer to the carbon star phases.}
\label{fmod}
\end{figure*}


The discussion presented in the following sections will be based on AGB evolution models
which include the description of dust formation in the circumstellar envelope.
The evolutionary sequences on which the present work is based are the same ones used in a 
series of papers published by our group to characterize the evolved stellar populations
of the Magellanic Clouds \citep{flavia14b, flavia15a, flavia15b} and Local Group 
galaxies \citep{flavia16, flavia18b, flavia19}. The interested reader can find in these
papers and in the recent review by \citet{karakas14} a thorough discussion of the AGB 
evolution of $1-8~M_{\odot}$ stars. Here we briefly recall the 
most relevant properties.

\subsection{The physical and chemical evolution of AGB stars}
\label{properties}
The left and middle panels of Fig.~\ref{fmod} show the evolution of the luminosity and of the $^{12}$C
surface mass fraction of stars of different mass. The grey parts of the tracks 
correspond to the C-star phase. Among the various chemical species we focus on $^{12}$C, 
because surface carbon is extremely sensitive to the efficiency of both hot bottom burning (hereinafter HBB, 
Renzini \& Voli 1981) and third dredge-up \citep[TDU,][]{iben74}. The models presented in 
Fig.~\ref{fmod} have $Z=8\times 10^{-3}$, the metallicity 
shared by the majority of the stars in the LMC, now evolving through the AGB phase \citep{harris09}. 
The current mass of the star is shown on the abscissa.

HBB consists in the activation of a series of p-capture reactions at the base of the
convective envelope, which modifies the relative distribution of the various
chemical species. The ignition of this process requires temperature at the bottom of
the surface convective zone of the order of $\sim 30$ MK, a condition that requires 
core masses $\sim 0.8~M_{\odot}$ \citep{ventura13}, reached only by stars of initial 
mass above
$3.5~M_{\odot}$\footnote{The minimum initial mass required to activate HBB during the
AGB evolution is indeed dependent on the metallicity of the star. It is $3.5~M_{\odot}$
for the $Z=8\times 10^{-3}$ chemistry used here, whereas for metal-poor stars with
$Z \leq 10^{-3}$ it is $\sim 3~M_{\odot}$.}. 

The ignition of HBB (see the tracks of $4, 5, 6~M_{\odot}$ models in Fig.~\ref{fmod})
leads to a fast rise in the luminosity of the star \citep{blocker91}, which increases
during the first part of the AGB evolution, until reaching a peak of the order
of $5\times 10^4 - 10^5~L_{\odot}$, which is higher the larger the 
initial mass of the star is; in the final AGB phases the luminosity diminishes,
because the general cooling of the external regions weakens (and eventually turns off)
HBB \citep{mazzitelli99}.

The activation of HBB favours the destruction of the surface carbon, 
exposed to proton fusion at the base of the envelope; this effect can be seen in 
the steep drop (by a factor $\sim 20$) in the surface $^{12}$C, clearly visible in middle 
panel of Fig.~\ref{fmod}. Depletion of surface carbon requires temperatures 
at the base of the envelope of the order of $\sim 40$ MK, which are reached by all the
stars experiencing HBB, independently of the chemical composition. Other nuclear 
reactions, e.g. proton captures by oxygen nuclei and the activation of the Ne-Na and
Mg-Al-Si nucleosynthesis, demand temperatures close to 100 MK, that are reached only
by AGB stars of low metallicity \citep[$Z \leq 10^{-3}$,][]{flavia18a}.
The tracks corresponding to the $1.25, 1.5, 2.5~M_{\odot}$ models in Fig.~\ref{fmod}
represent low-mass AGB stars, not experiencing HBB. In these stars the luminosity rises
during the AGB phase, owing to the increase in core mass, until reaching a final value,
slightly below $\sim 10^4~L_{\odot}$ \citep{boothroyd88}. 

The only mechanism able to alter the surface chemical composition of these stars is TDU, 
which provokes a gradual increase in the surface carbon, evident in the middle panel 
of Fig.~\ref{fmod}. Repeated TDU events lead to the formation of carbon stars, with a
surface C$/$O ratio above unity. This condition is reached by stars with initial mass
$1 < M/M_{\odot} < 3$\footnote{The lower limit in mass is slightly dependent on
metallicity. Indeed in metal-poor stars the achievement of the C-star condition is easier,
because the lower oxygen; therefore, the lower mass
threshold required to reach the C-star phase during the AGB evolution is smaller.}.
The largest values of surface carbon mass fraction, slightly above $\sim 1\%$, are reached by the 
stars of initial mass $\sim 2.5-3~M_{\odot}$, that are exposed to a higher number of TDU 
episodes before they loose the external envelope, compared to their lower mass counterparts
\citep{karakas10, karakas18}.  

On the physical side, the enrichment in carbon content in the envelope determines a significant
increase in the surface opacities \citep{marigo02}, which favours the expansion of the 
external regions of the star and the increase in the mass loss rate, that exceeds
$\sim 10^{-5}~M_{\odot}/$yr \citep{vm09, vm10}. The
stars of initial mass below $2~M_{\odot}$ become C-stars after a significant fraction of the
envelope was lost during the previous AGB phases; for the reasons given above the mass loss 
rates experienced after the C-star stage is reached are so large that the residual
envelope is lost before further thermal pulses (TPs) are experienced.

A general behaviour of AGB stars is that the loss of the external mantle is accompanied 
by the cooling of the external regions, which makes the stars readjust on a more and 
more expanded configuration \citep{vw93}. As TP-AGB stars are pulsating stars, this reflects 
into a gradual increase of the pulsation period (P) of the star. Massive AGB stars are 
expected to reach extremely long periods, in the range 1000-2000d, during the peak of the 
HBB activity; in the finale phases, after most of the envelope is lost and the star starts 
to contract, the pulsation periods decrease.

\subsection{Dust production in AGB stars}
\label{dustproduction}
In the schematization adopted here dust formation
occurs while the gas is expanding away from the central star, in the form of wind.
The key-factor affecting the mineralogy of the dust formed is the C$/$O ratio, owing to
the high stability of the CO molecule \citep{sharp90}. In oxygen-rich stars the formation of 
silicates, alumina dust and solid iron occurs, whereas in the envelope of carbon stars the 
main dust species formed are solid carbon, SiC and solid iron \citep{fg06}. 

The calculation of the extinction coefficients, required to determine the acceleration
of the wind, is based on the following set of optical constants: silicates
\citep{ossenkopf92}; alumina dust \citep{begemann97}; iron \citep{ordal88}, solid carbon
\citep{hanner88}; silicon carbide \citep{sic88}.

The dust mass loss rates ($\dot M_d$) for the AGB models discussed so far are shown in 
the right panel of Fig.~\ref{fmod}. These results are thoroughly discussed in
\citet{flavia15a}. The phases with the largest $\dot M_d$ are those
showing up the highest degree of obscuration, thus the largest IR excess.

Massive AGB stars never reach the C-star stage, thus they do not produce any carbon dust.
The $\dot M_d$ of these objects is mostly determined by the values of the gas mass loss 
rate, which reflects the run of the luminosity\footnote{The increase in the mass loss rate 
with the luminosity, which is general, 
is particularly steep in the present models, that adopt the treatment of mass loss by 
\citet{blocker91}.}. This behaviour is clear when comparing the evolution of luminosity and 
dust mass loss rate in $M \geq 4~M_{\odot}$ stars, shown in the left and right panels of 
Fig.~\ref{fmod}.

In low-mass stars the mineralogy of the dust formed consists in silicates, alumina 
dust and iron, during the first part of the AGB evolution, then changes to carbonaceous 
species, when they become carbon stars. As shown in Fig.~\ref{fmod}, the achievement of 
the C-star stage provokes a significant increase in $\dot M_d$. 

We now focus on the dust produced during the O-rich phase, which is most relevant to the
present work. Most of the dust from M-stars is produced by $M \geq 4~M_{\odot}$ stars,
during the phases that follow the ignition of HBB. Inspection of the right
 panel of Fig.~\ref{fmod} reveals that these stars experience an initial AGB phase,
with poor dust production, followed by phases characterized by the presence of large
quantities of dust in the circumstellar envelope, with 
$\dot M_d \sim 10^{-8}-10^{-7}~M_{\odot}/$yr. Low-mass objects evolve as M-stars
during the initial part of the AGB phase, before they turn to C-stars. The dust
produced by these objects during the O-rich phase is negligible, because their mass loss 
rate, below $10^{-6}~M_{\odot}/$yr, is too small to allow formation of dust in
meaningful quantities. An exception to this is represented by low-mass stars of initial 
mass around $1-1.5~M_{\odot}$. These stars reach the C-star phase only in the very final 
AGB stages, when a significant fraction of the envelope is lost. During the end of the 
O-rich phase, before turning to C-stars, they evolve at cool effective temperatures and 
their mass loss rate reaches $\sim 10^{-6}~M_{\odot}/$yr: these conditions favour the 
formation of small, but not negligible, amount of silicates and alumina dust.

\begin{figure*}
\begin{minipage}{0.48\textwidth}
\resizebox{1.\hsize}{!}{\includegraphics{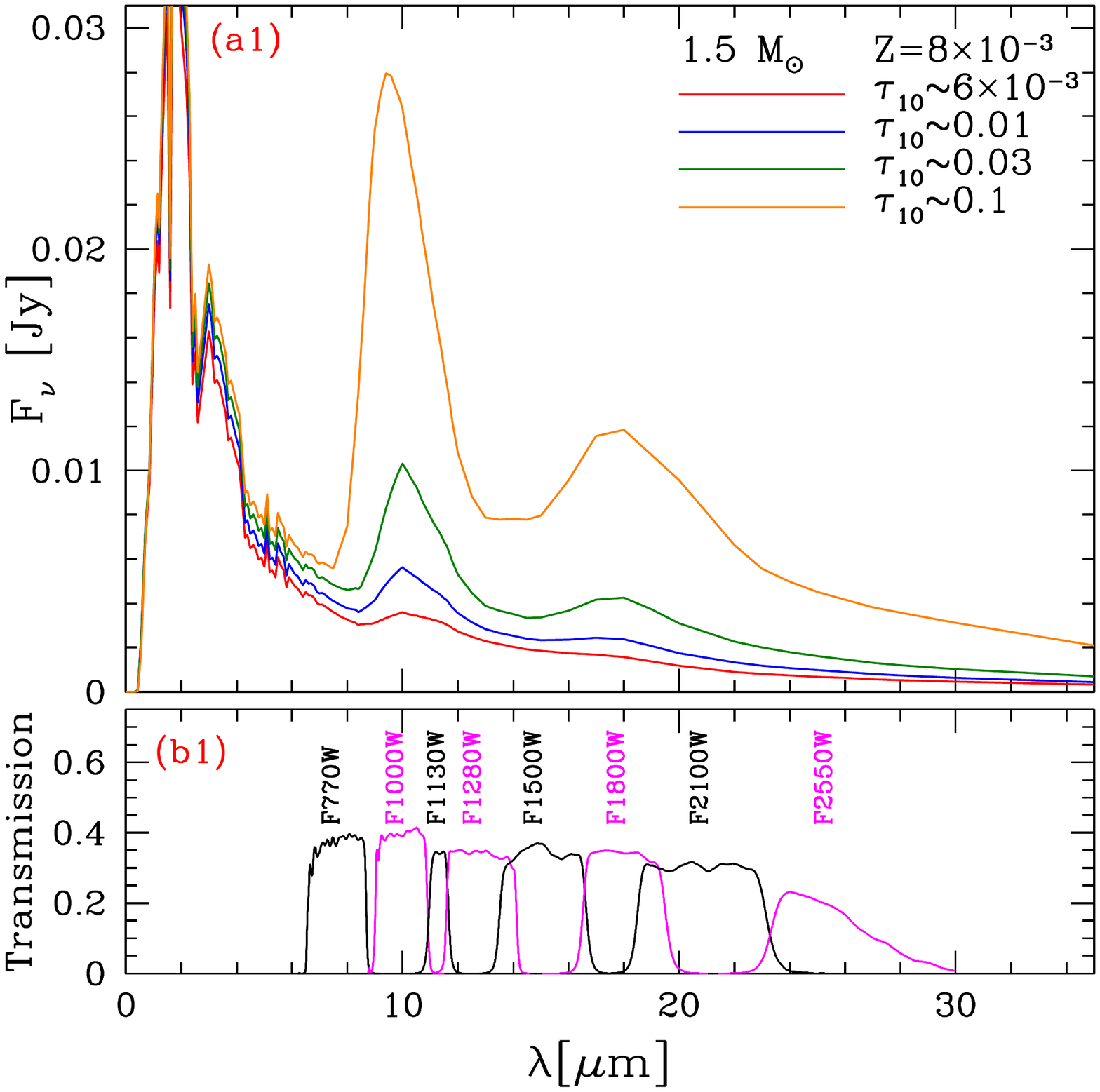}}
\end{minipage}
\begin{minipage}{0.48\textwidth}
\resizebox{1.\hsize}{!}{\includegraphics{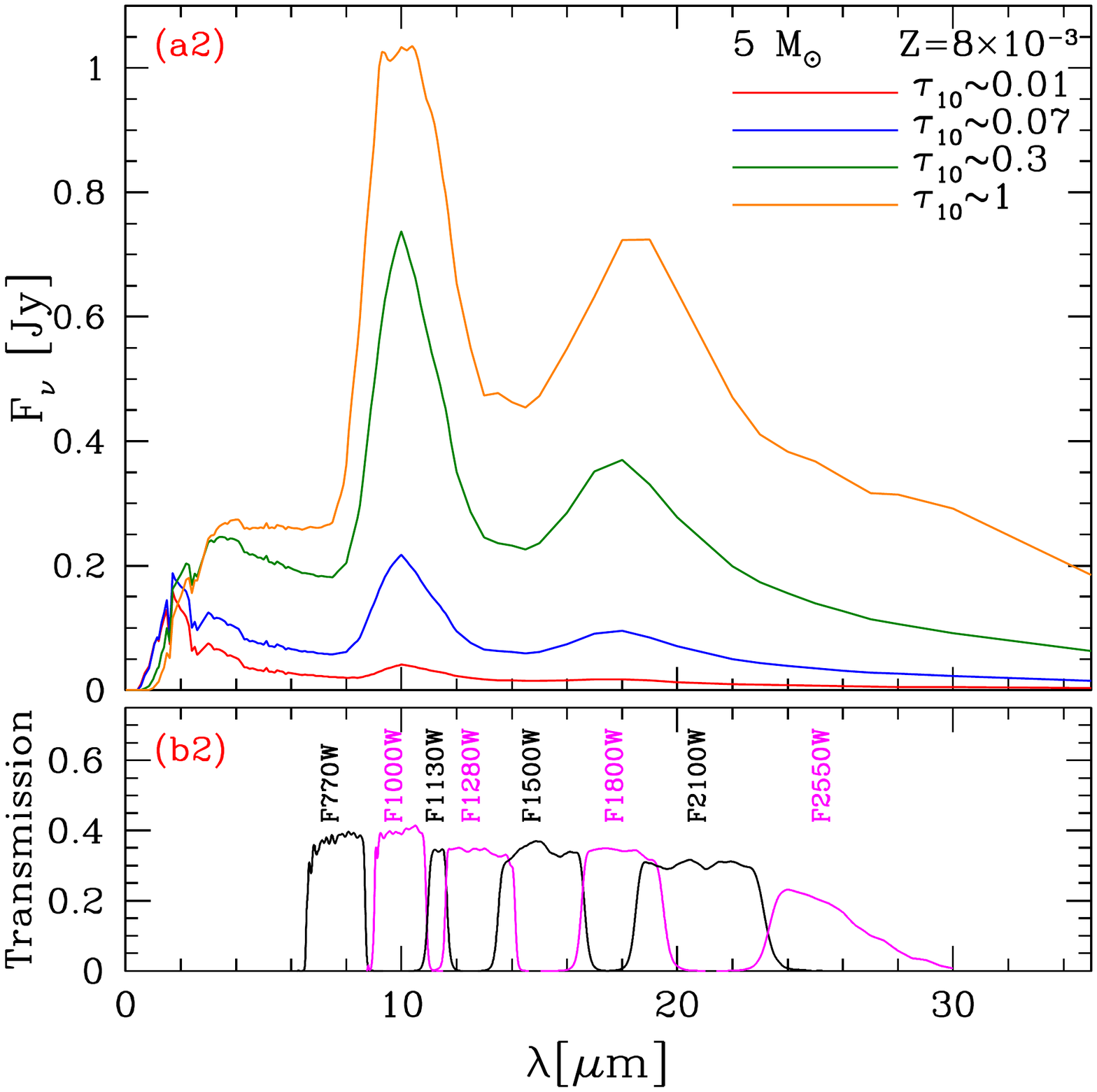}}
\end{minipage}
\vskip-60pt
\caption{The expected variation of the SED during 4 evolutionary phases
of the AGB evolution of a star of initial mass $1.5~M_{\odot}$ (a1) and
$5~M_{\odot}$ (a2). The 4 phases are characterized by different degrees of obscuration
of the circumstellar envelope, here represented by the values of $\tau_{10}$, indicated 
in the two panels. Fluxes are calibrated to the distance of the LMC \citep{feast99}. 
Panels b1 and b2 show the transmission curves of MIRI filters in el. sec$^{-1}$.}
\label{fspet}
\end{figure*}

\subsection{The evolution of the spectral energy distribution of M-type AGB stars}
\label{spettro}
The results from stellar evolution modelling coupled with the description of dust 
formation in the wind allows the determination of the change in the spectral energy 
distribution (SED) of the star during the AGB phase. To this aim, we selected some 
representative points along the individual sequences (tipically $\sim 20$ between two 
successive TPs) and used the code DUSTY \citep{nenkova99} to calculate the shape of the 
expected emission. The input needed to this computation are the temperature of the 
condensation region, luminosity, effective temperature of the star, the size of the dust 
grains formed and the optical depth, which in our case is 
assumed at the wavelength $\lambda=10 \mu m$.

From the discussion in the previous section we know that in the context of M stars 
significant dust quantities are produced by $M \leq 2~M_{\odot}$ stars, in the phases 
previous to the achievement of the C-star stage, and by massive AGB stars, after 
the beginning of HBB. We discuss these two cases below, separately. 

The former
evolve as M stars for most of their AGB life, until they become carbon stars. 
$\dot M_d$ becomes higher and higher during the AGB phase (see the blue and magenta tracks,
corresponding to the $1.5~M_{\odot}$ and $1.25~M_{\odot}$ models, in the right 
panel of Fig.~\ref{fmod}), because the gradual loss of the envelope and the general 
cooling of the external regions favour the increase in the mass loss rate. The 
degree of obscuration in the SED of these objects thus increases during the M-stage. This is
shown in panel a1 of Fig.~\ref{fspet}, that shows the expected evolution of
the SED of a star of initial mass $1.5~M_{\odot}$; the different lines correspond to
four evolutionary stages, from the initial AGB phases (red line), when dust 
formation is inefficient, to the phases immediately before the achievement of the C-star phase 
(orange track), when the optical depth grows to $\tau_{10} \sim 0.1$. The increase in the 
amount of dust formed makes the SED of the star to exhibit two prominent features, at 
$9.7~\mu$m and $18.7~\mu$m, associated to the presence of silicates. Furthermore, the SED 
in the whole mid-IR region of the spectrum is lifted.

The stars that reach the highest $\dot M_d$ during the M-stage are those
that experience HBB (see red, black and magenta lines in the right panel of 
Fig.~\ref{fmod}), the progeny of $M \geq 4~M_{\odot}$ stars. As shown in Fig.~\ref{fmod},   
$\dot M_d$ of this class of objects increases during the first part of the AGB evolution, 
until the maximum luminosity and mass loss rate is reached, then decreases when the 
efficiency of HBB starts to diminish. According to our modelling the largest optical 
depths attained are of the order of $\tau_{10} \sim 1$ \citep{flavia15a}.

Panel a2 of Fig.~\ref{fspet} shows the expected evolution of the SED of a $5~M_{\odot}$ 
star, from the beginning of the AGB phase (red line), until the point of highest 
obscuration, reached in conjunction with the largest luminosity (orange). 

The changes in the SED of these two classes of objects allow the determination of
the general obscuration patterns traced by these stars in the different observational planes,
that are useful to characterize the evolved M stars in the LMC and in other galaxies.

\subsection{Post-AGB evolution modelling}
\label{post}
A few sources belonging to the sample published by \citet{jones12} show the presence of 
cool dust in the circumstellar envelope, suggesting that dust production has stopped and 
that the dust currently observed was produced during earlier evolutionary phases. 
The SED of these stars present a peak in the optical that rules out the possibility 
that they are AGB stars. Based on these factors, \citet{ester19b} suggested that 
these stars have left the AGB and are evolving through the post-AGB phase.

To further investigate and interpret these objects, we extended the evolutionary
computations, so far limited to the AGB stage\footnote{When a few tenths of solar masses were left 
in the envelope.}, to the post-AGB phase. These sources, investigated by \citet{ester19b}
and further discussed in detail in Sect.~\ref{dualstars}, have luminosities below 
$10^4~L_{\odot}$, compatible with the evolution of stars of initial masses below 
$2~M_{\odot}$. Therefore, we decided to focus on the post-AGB phase of 
$M \leq 2~M_{\odot}$ stars, with metallicity $Z=8\times 10^{-3}$.

\begin{table*}
\caption{Main physical and chemical properties of the post-AGB models discussed
in the text. Cols. 1-3 indicate the initial mass of the star, the mass at the
beginning of the AGB phase and the final mass. The time scales reported in col. 4 and
5 indicate, respectively, the age of the star and the crossing time, defined as
the time interval from the point when the mass of the envelope drops below $1\%$
of the stellar mass and the time when the effective temperature becomes 
$\log T_{\rm eff}=3.85$.
The last four cols. report the surface chemical composition, in terms of the mass fractions
of helium, carbon, nitrogen and oxygen.}                                       
\begin{tabular}{c c c c c c c c c}        
\hline
$M/M_{\odot}$  & $M_{AGB}/M_{\odot}$  & $M_f/M_{\odot}$  &  $\tau_{ev}$ (Gyr)& $\tau_{tr}$ (Kyr)  &  X(He)  &  X(C)  &  X(N)  &  X(O)   \\
\hline
1.00  &  0.75  &  0.552  &  9.23  &  3.8  &  0.281  &  8.86e-4  &  4.72e-4  &  4.16e-3  \\
1.10  &  0.90  &  0.575  &  6.52  &  8.8  &  0.281  &  9.74e-4  &  3.08e-4  &  4.24e-3  \\
1.25  &  1.10  &  0.596  &  4.22  &  3.3  &  0.290  &  8.59e-4  &  3.24e-4  &  4.66e-3  \\ 
1.40  &  1.25  &  0.590  &  2.91  &  4.2  &  0.284  &  3.91e-3  &  3.02e-4  &  4.39e-3  \\
1.60  &  1.50  &  0.602  &  1.89  &  3.4  &  0.284  &  3.86e-3  &  2.99e-4  &  4.40e-3  \\ 
1.75  &  1.75  &  0.622  &  1.42  &  2.9  &  0.285  &  4.84e-3  &  3.14e-4  &  4.62e-3  \\
2.00  &  2.00  &  0.617  &  0.97  &  3.0  &  0.286  &  6.21e-3  &  2.97e-4  &  4.68e-3  \\
\hline     
\label{tabpostagb}
\end{tabular}
\end{table*}

The present post-AGB computations were self-consistently resumed from the $Z=8\times 10^{-3}$
AGB models used by \citet{flavia15a}.
For the stars not reaching the C-star stage we modeled mass loss according to eq.~6 in
\citet{marcelo16}. For C-stars we chose,
somewhat arbitrarily, to keep the description by \citet{wachter02, wachter08}. This choice
will likely affect the timescale of the post-AGB evolution, but is not relevant for the
determination of the excursion of the evolutionary track on the HR diagram.

The main properties of the post-AGB models are reported in Table.~\ref{tabpostagb}.
As discussed by \citet{marcelo16}, the definition of the start of the post-AGB phase is 
not rigorous, considering that the behaviour of the stars, particularly the beginning
of the excursion of the evolutionary track to the blue, is dependent on the core mass. We 
assume that the AGB evolution ends when the mass of the envelope drops below
$1\%$ of the mass of the star and chose the point when the effective temperature is
$\log T_{\rm eff}=3.85$ as the intermediate stage between the post-AGB phase and the
PN evolution. The time interval between these two evolutionary stages, proposed
by \citet{marcelo16} as indicator of the transition time scale from the AGB to the
post-AGB phase, is reported on column~5 of Table~\ref{tabpostagb}.

\begin{figure}
\begin{minipage}{0.48\textwidth}
\resizebox{1.\hsize}{!}{\includegraphics{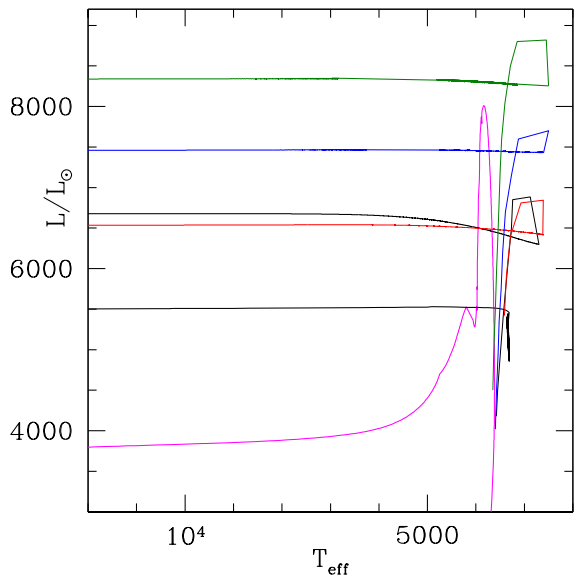}}
\end{minipage}
\vskip-60pt
\caption{The path traced in the HR diagram by stars of initial mass
$1~M_{\odot}$ (magenta line), $1.1~M_{\odot}$ (black), $1.25~M_{\odot}$ (red), 
$1.4~M_{\odot}$ (black), $1.6~M_{\odot}$ (blue), $1.75~M_{\odot}$ (green), 
during the latest AGB phases and the post-AGB phase.}
\label{fpagb}
\end{figure}

The path traced by the evolutionary tracks in the HR diagram of the models reported in 
Table~\ref{tabpostagb} are shown in Fig.~\ref{fpagb}. Since
the sources that we want to discuss here have $T_{\rm eff} < 10^4$K, we 
restrict our attention to effective temperatures below $1.2\times 10^4$ K. For
each mass we show a few points regarding the very final AGB phases and the 
whole post-AGB evolution.

These results will be used later to discuss a few stars in the sample that
are most likely evolving through the post-AGB phase.

\section{The LMC sample}
In this paper we use the LMC sample of 71 O-rich AGB spectra\footnote{http://irsa.ipac.caltech.edu/data/SPITZER/SAGE}, 
observed using {\itshape Spitzer} IRS \citep{jones12} and classified according to the decision-tree scheme 
proposed by \citet{woods11}. The sample includes point sources from the SAGE-Spec legacy 
survey of the LMC \citep{kemper10}, a spectroscopic follow-up to the SAGE-LMC 
project \citep{meixner06} and has been supplemented from archival {\itshape Spitzer} IRS observation within 
the SAGE-LMC footprint \citep{woods11}. The detailed description of the original target 
selection, the observing strategy and the techniques used in the data reduction for the 
SAGE-Spec legacy programme are discussed in \citet{kemper10}.\\
To study the distribution on the observational planes built from MIRI filters, we use the 
mid-IR magnitudes ([F770W], [F1000W], [F1130W], [F1280W], [F1500W], [F1800W], [F2100W], 
[F2550W]) calculated by \citet{jones17}, who integrated the {\itshape Spitzer} IRS spectra of each source 
over the MIRI spectral response (see panels b1 and b2 in Fig.~\ref{fspet}). The fluxes for the F560W filter are not available since 
the {\itshape Spitzer} IRS spectra cover the wavelength range $5.3-8~\mu$m.
For all the sources the associated broadband photometry, including optical $UBVI$ 
photometry from the Magellanic Clouds Photometric Survey \citep{zaritsky04}, Two Micron 
All Sky Survey (2MASS) $JHK_{S}$ photometry \citep{skrutskie06}, mid-IR photometry from 
Infrared Array Camera (IRAC 3.6, 4.5, 5.8, 8.0$~\mu m$) and Multi-Band Imaging Photometer 
for {\itshape Spitzer} (MIPS 24$~\mu m$), was compiled from the SAGE catalogue \citep{meixner06}. 
For 23 of these objects there is no 
spectral coverage by the {\itshape Spitzer}-IRS at $\lambda > 14.2~\mu$m, so mid-IR photometry beyond 
this wavelength is not available.\\
The list of the stars analyzed in this work is reported in Table 2. When available, in 
column 3 we report the primary periods obtained by the OGLE collaboration \citep{soszy09}, 
that phases well the temporal sequence of observations.

\begin{figure*}
\begin{minipage}{0.48\textwidth}
\resizebox{1.\hsize}{!}{\includegraphics{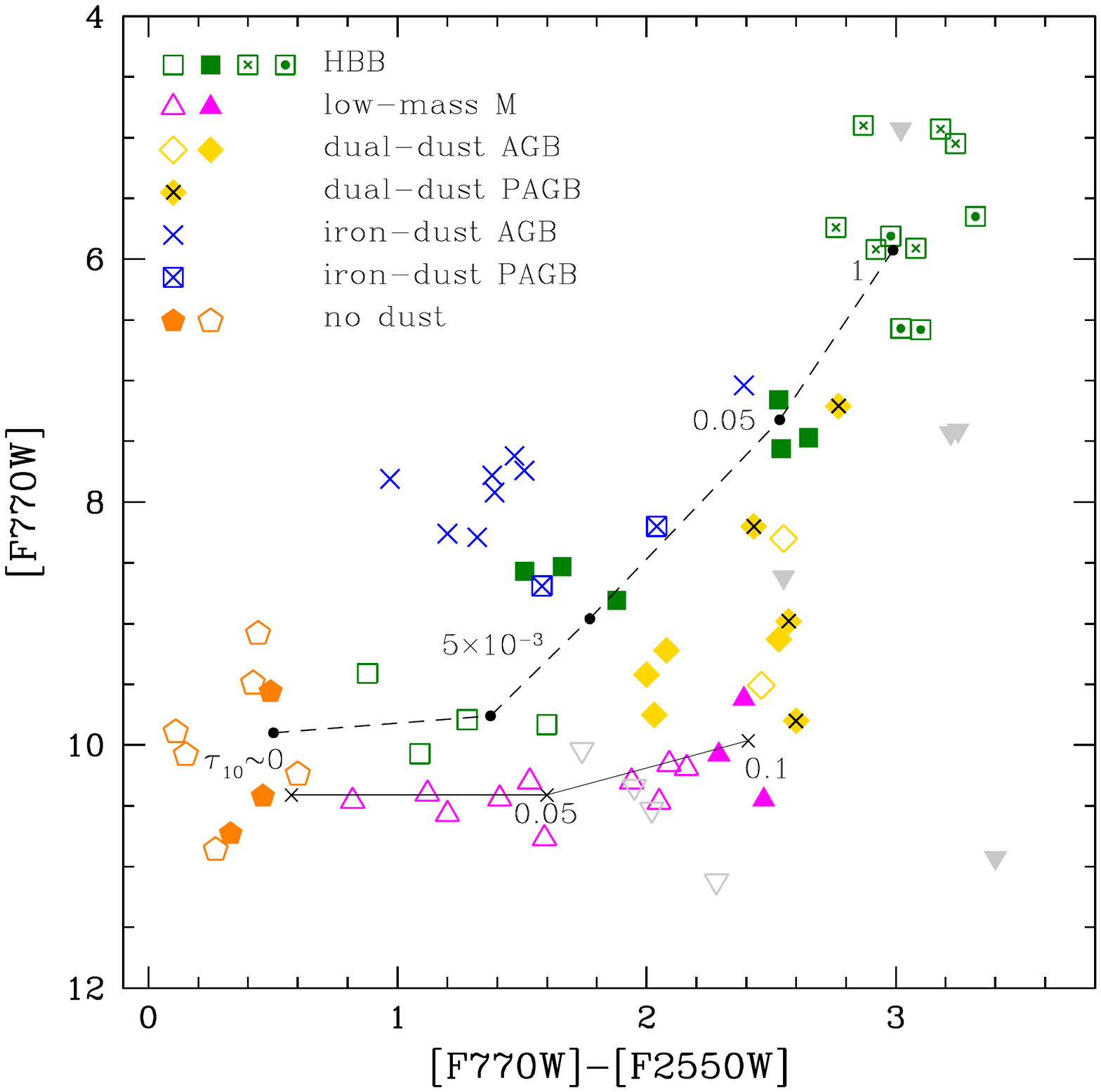}}
\end{minipage}
\begin{minipage}{0.48\textwidth}
\resizebox{1.\hsize}{!}{\includegraphics{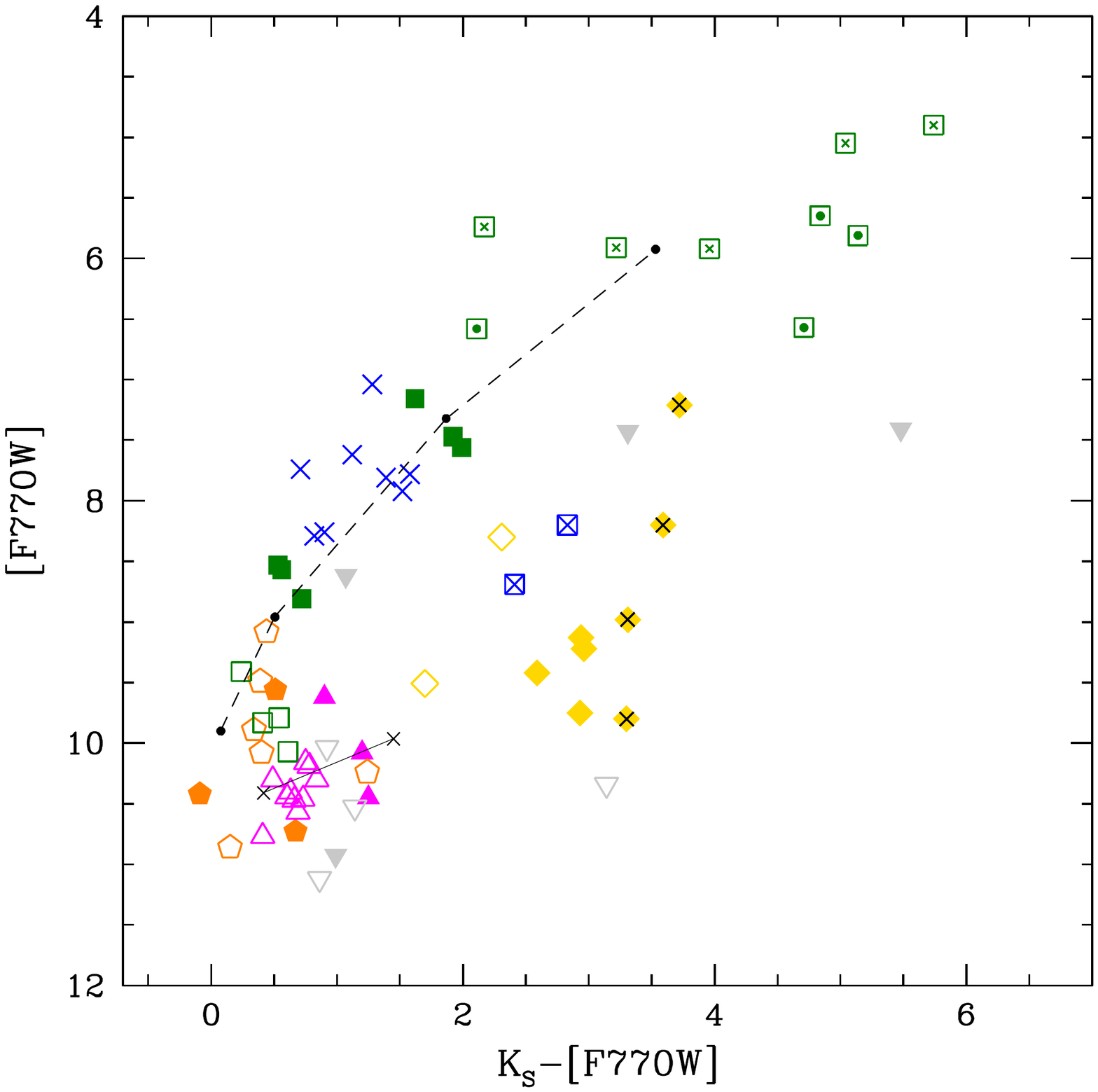}}
\end{minipage}
\vskip-80pt
\begin{minipage}{0.48\textwidth}
\resizebox{1.\hsize}{!}{\includegraphics{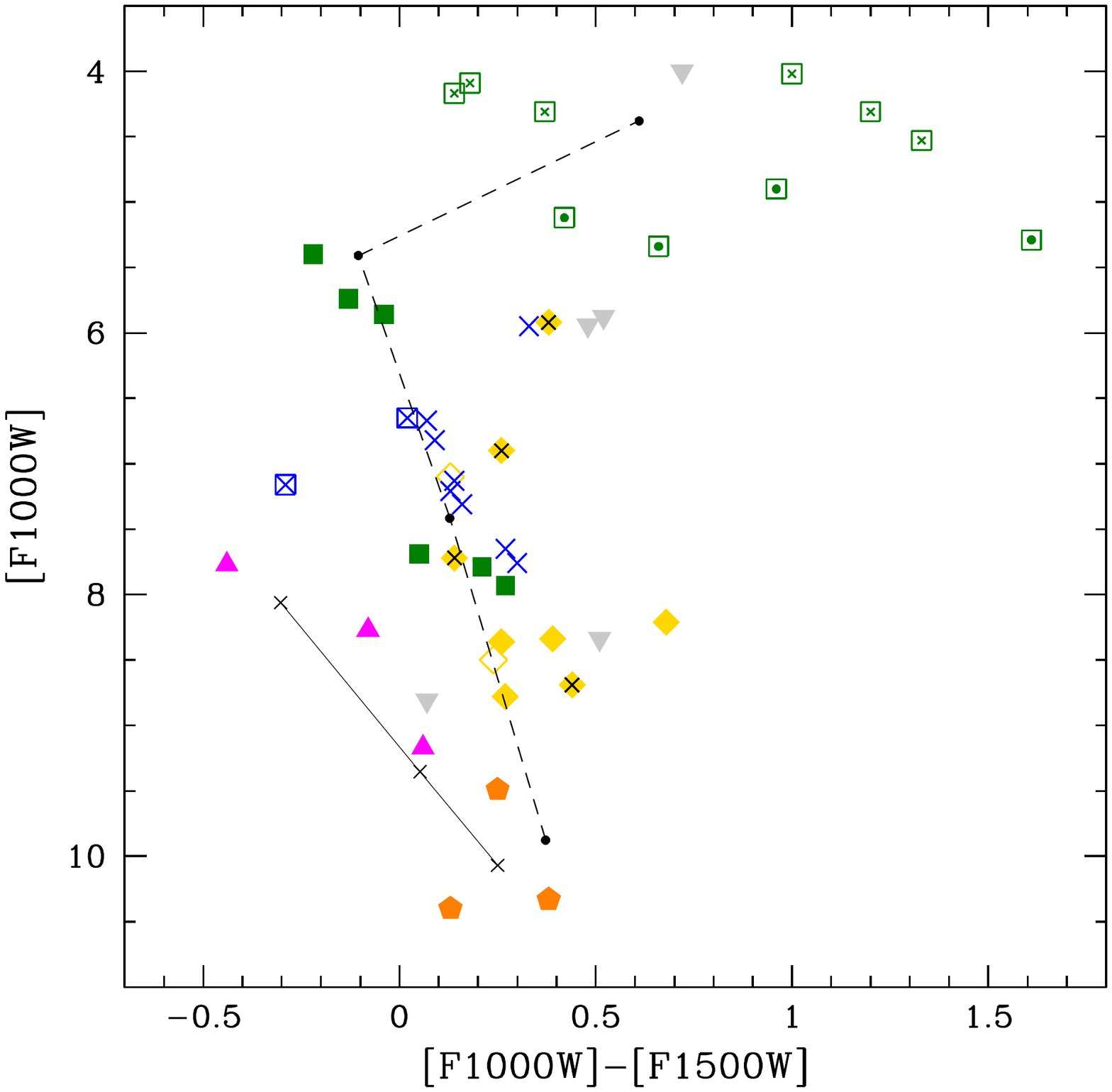}}
\end{minipage}
\begin{minipage}{0.48\textwidth}
\resizebox{1.\hsize}{!}{\includegraphics{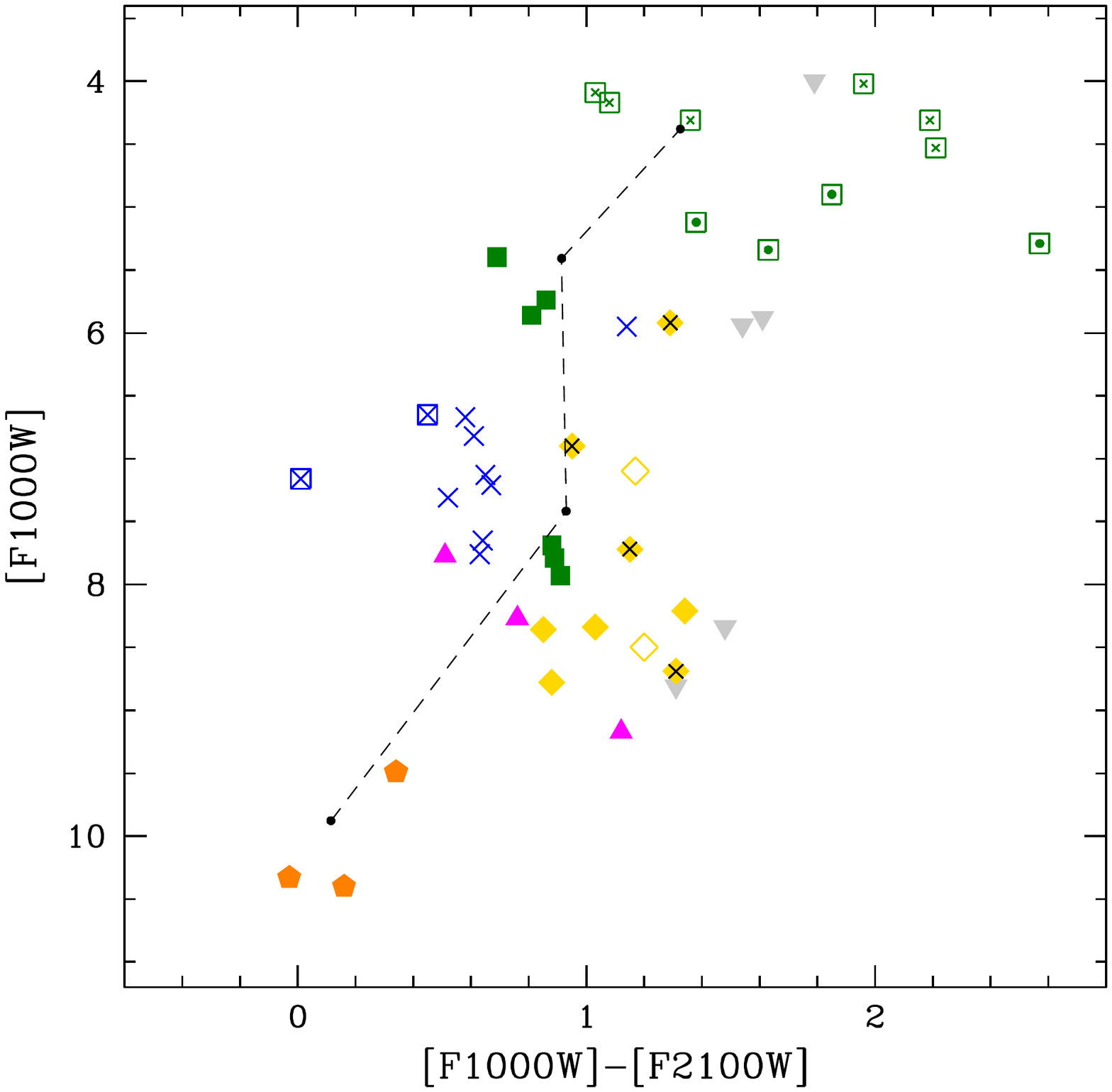}}
\end{minipage}
\vskip-60pt
\caption{The distribution of the LMC stars belonging to the sample studied by \citet{jones12, jones14, jones17}
in various observational planes, built with different combinations of the filters of the 
MIRI camera, mounted onboard of the {\itshape JWST} and 2MASS $K_S$ filter. The values of the 
magnitudes have been obtained by convolving {\itshape Spitzer} IRS data with the transmission curves 
of the various filters. For the stars indicated with open points, the IRS data were 
truncated to $\lambda \sim 14.2~\mu$m; in these cases we used the MIPS $[24]$ magnitude,
when available, as a rough estimate of [F2550W], as the difference between them is only a few hundredth of magnitudes. The different symbols refer to the various
classes of objects, according to the legend reported in the top, left panel. Among these, grey triangles indicate 
sources for which we could not provide a reliable interpretation on the basis of their SED. The solid and
dashed lines indicate the obscuration sequence of low-mass, oxygen-rich AGB stars and
of massive AGBs experiencing HBB, respectively (see Sect.~\ref{lowmass} and \ref{hbbstars} for details). 
The black points along the two sequences refer to typical values of the optical depth $\tau_{10}$, which 
are indicated in the top, left panel.}
\label{fcmd}
\end{figure*}

\begin{figure}
\begin{minipage}{0.48\textwidth}
\resizebox{1.\hsize}{!}{\includegraphics{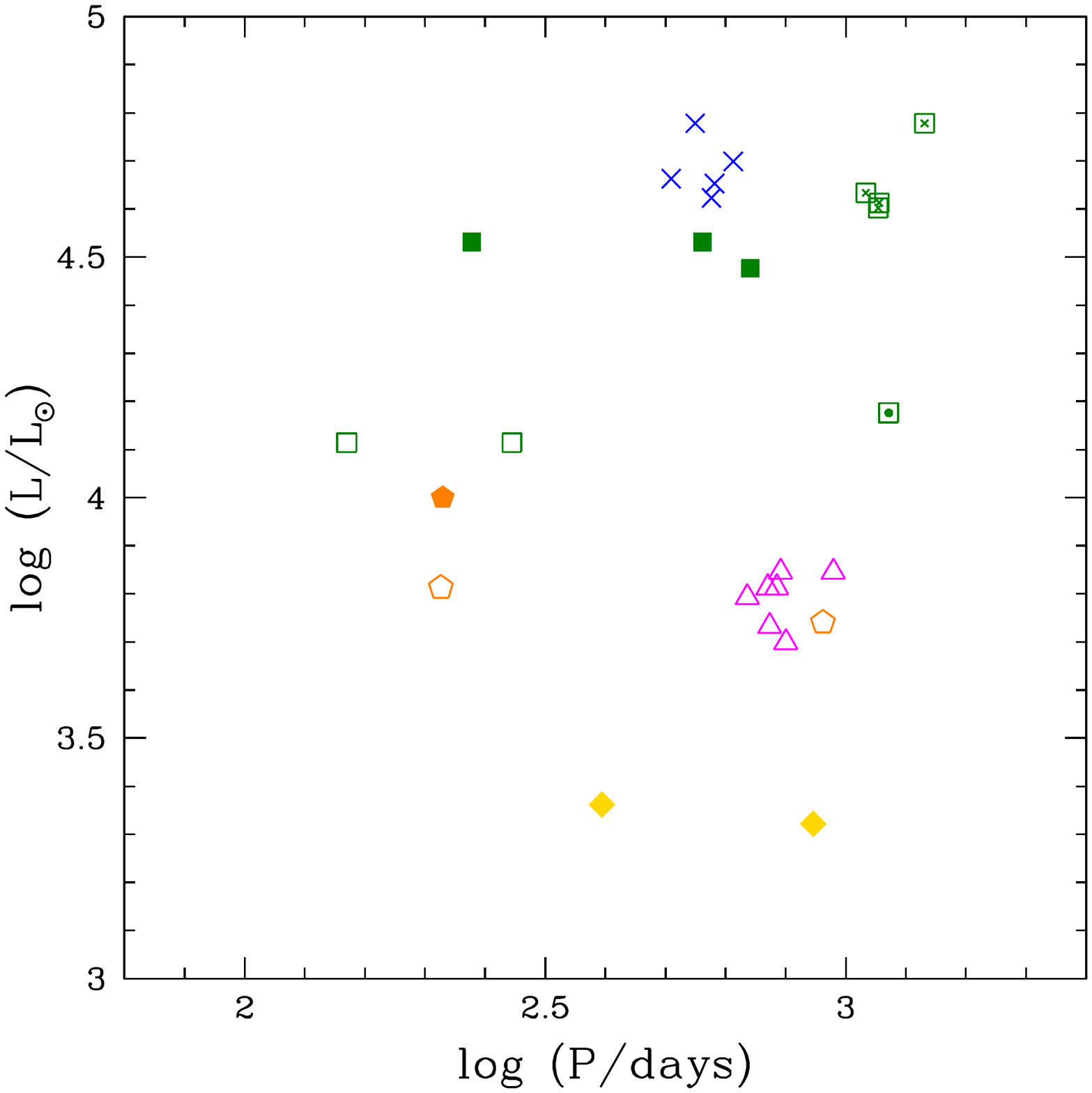}}
\end{minipage}
\vskip-60pt
\caption{Period - Luminosity diagram of the LMC M-type stars in the sample for which OGLE 
periods are available. Luminosities are those reported in Table 2. Symbols are the same 
used in Fig.~\ref{fcmd}.}
\label{pl}
\end{figure}

\section{M-type AGB stars: a {\itshape JWST} perspective}
Fig.~\ref{fcmd} shows the distributions of the stars studied by \citet{jones12, jones14, jones17} 
in four colour-magnitude planes obtained by means of the fluxes in some of the {\itshape JWST} 
filters. Three out of the four planes, built with the MIRI filters, were proposed by \citet{jones17} 
to distinguish different classes of objects and identify C- and M-type AGB stars. 
In addition, we discuss the ($K_{S}$-[F770W],[F770W])\footnote{We expect to obtain similar 
trends when using the NIRCam F210M filter.} plane to explore the possible use of 
the combination of near-IR and mid-IR photometry to deduce the properties of the
stars observed.  

Our approach is the following. We first characterize the individual sources based on 
the comparison between their position in the afore mentioned planes and the path traced by 
the evolutionary tracks, calculated by convolving the synthetic SEDs, discussed in 
Sect.~\ref{spettro}, with the transmission curves of the MIRI filters, shown in panels 
b1 and b2 of Fig.~\ref{fspet}, and of 2MASS K$_S$ filter. On this regard, we believe 
important to underline that low-mass stars and massive AGBs define two distinct obscuration 
patterns, where the colors and magnitudes are primarily determined by the luminosity and 
the degree of obscuration, i.e. the optical depth. We will return to this point in 
Sect.~\ref{lowmass} and  \ref{hbbstars}.  We further define the details of the dust 
mineralogy by the tight comparison between the IRS spectra and the synthetic SEDs, 
corresponding to the individual points along the tracks.

Fig.~\ref{pl} shows the distribution in the Period - Luminosity (PL) diagram of the stars 
in our sample for which OGLE periods are available. In some cases this plane will help 
us to better characterize the sources.

In the following we discuss different groups of stars, separated according to their
degree of obscuration and/or peculiar features present in the SED, in turn connected
with the mineralogy of the dust in the circumstellar envelope.

\begin{figure*}
\begin{minipage}{0.48\textwidth}
\resizebox{1.\hsize}{!}{\includegraphics{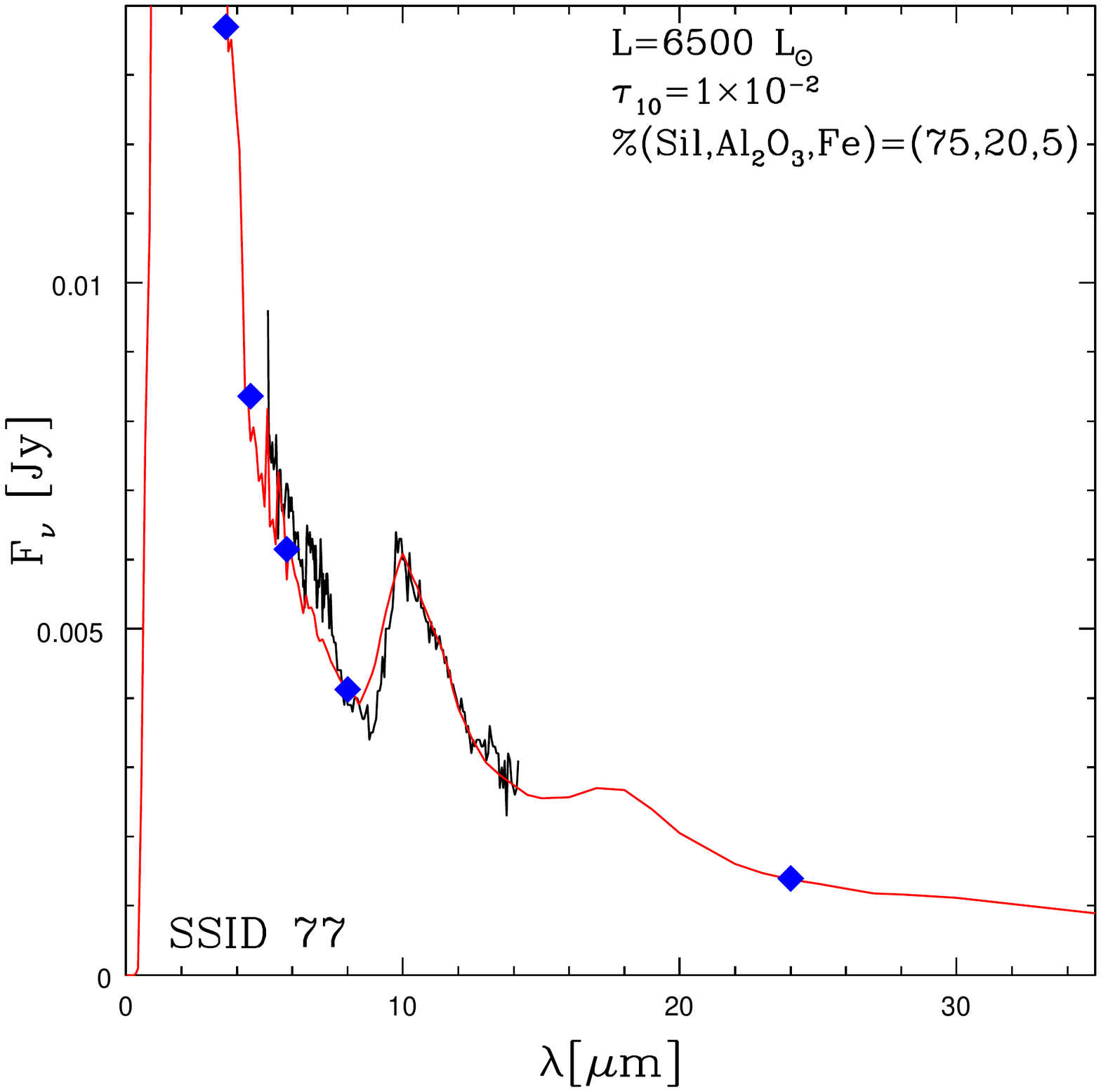}}
\end{minipage}
\begin{minipage}{0.48\textwidth}
\resizebox{1.\hsize}{!}{\includegraphics{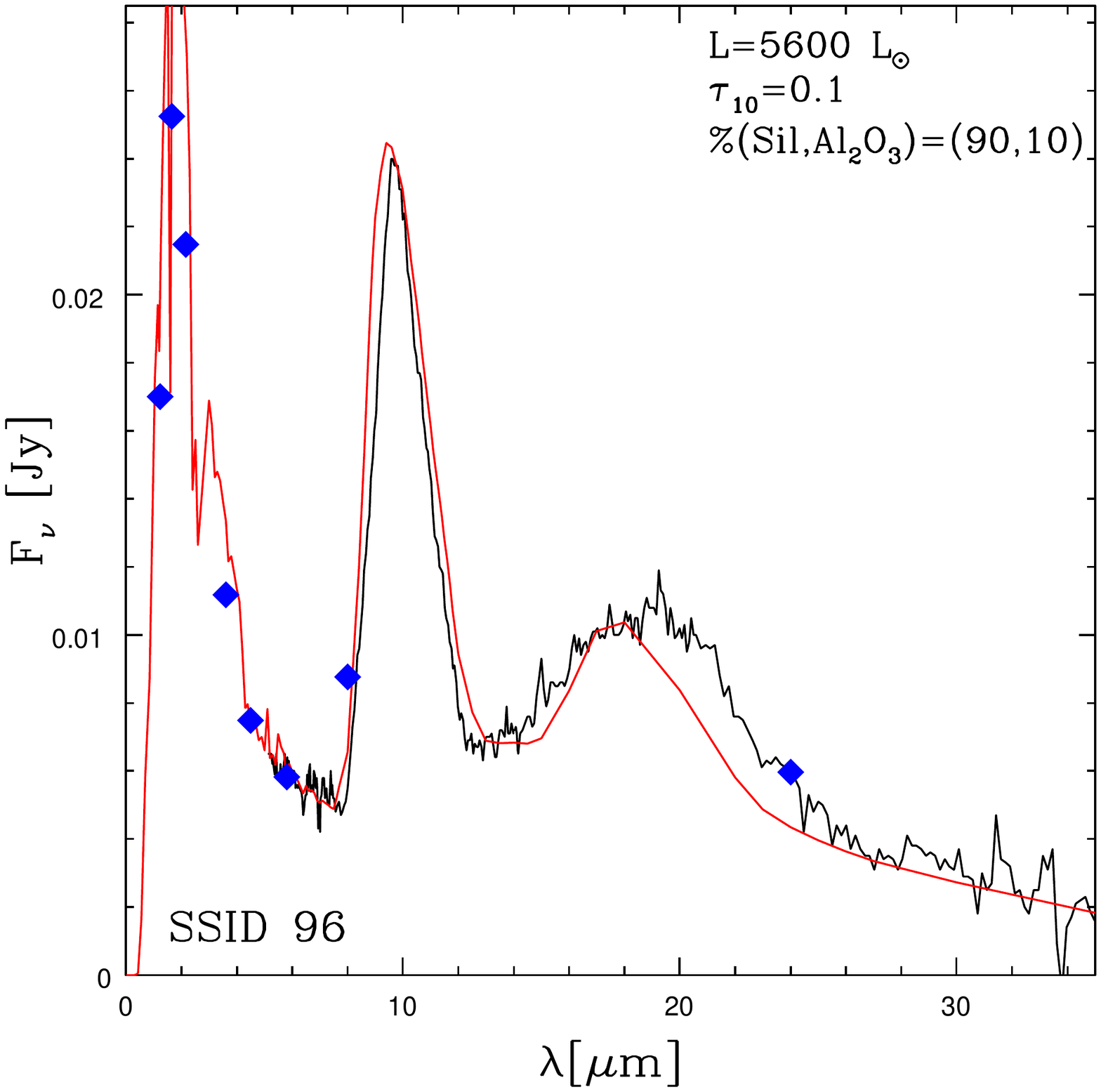}}
\end{minipage}
\vskip-60pt
\caption{The {\itshape Spitzer} IRS spectra (black line) of two stars, selected among those interpreted
as low-mass AGBs, taken along the obscuration sequence of low-mass stars
shown in Fig.~\ref{fcmd}. The photometry is indicated with blue diamonds.
The best fit, indicated with the red line, is obtained by assuming the luminosity, 
dust composition and optical depth indicated in the two panels.}
\label{flowm}
\end{figure*}

\subsection{Scarcely obscured, no-dusty stars}
\label{nodust}
The stars indicated with orange pentagons in Fig.~\ref{fcmd} show no trace of dust in 
their surroundings. The SED of these objects can be safely 
reproduced by assuming optical depths close to zero, indicating a negligible degree of
obscuration. Based on the luminosities obtained by SED fitting, we deduce that this 
group of sources is mainly composed by low-mass stars, that have not reached the C-star 
stage. In the ([F770W]-[F2550W], [F770W]) plane they populate the region within 
$0 <$ [F770W]-[F2550W] $< 0.7$, separated from the other sources.

In the ([F1000W]-[F1500W],[F1000W]) and ([F1000W]-[F2100W],[F1000W]) planes these stars 
can be identified by means of the low [F1000W] fluxes, consistent with the scarce presence 
of silicates in the circumstellar envelope, that prevents the formation of the $9.7~\mu$m 
feature. 

In the ($K_{S}$-[F770W], [F770W]) plane their identification is more tricky, because they 
mix with obscured stars with optical depths $0.005<\tau_{10}<0.1$. This is because when $0.005 < \tau_{10} < 0.1$:
a) the formation of silicates increases the flux in the spectral region around $10~\mu$m, 
provoking at the same time a dip around $\sim 8~\mu$m, that decreases the [F770W] flux 
(see e.g. the blue line in the right panel of Fig.~\ref{fspet}); b) there is
no significant shift of the SED towards mid-IR wavelengths, thus the $K_{S}$ flux is
only scarcely affected by dust reprocessing. Therefore, the stars with a small degree of 
obscuration are characterized by $K_{S}$-[F770W] colours and [F770W] magnitudes not 
significantly different from those exhibited by the unobscured stars discussed here.

In the planes shown in the top panels of Fig.~\ref{fcmd} the stars with no
dust span a [F770W] range of approximately 2 mag. This reflects the different
luminosities, which we find (by SED fitting) to be in the range $5\times 10^{3}-3\times 10^{4}~L_{\odot}$, 
consistent with the heterogeneity in the mass and formation epoch of the progenitors of 
these sources. This same analysis cannot be done in the planes shown in the bottom of the 
figure, because the lack of the {\itshape Spitzer} IRS data at wavelengths $\lambda > 14.2~\mu$m 
prevents the determination of [F1000W]-[F1500W] and [F1000W]-[F2100W] for the majority of 
these objects (those indicated with open pentagons in the top panels).

\subsection{Low-mass dusty M-type AGB stars}
\label{lowmass}
The magenta triangles in Fig.~\ref{fcmd} indicate stars that we interpret as low-mass dusty stars.
The evolutionary tracks of these objects in the different observational planes overlap substantially,
 mainly because of the similar luminosities, favoured by the occurrence of core electron degeneracy
for $M\leq2M_{\odot}$ progenitors. 

The redwards extension of the tracks depends on both the initial mass and the metallicity. 
Stars of initial mass $M\lesssim 1.5M_{\odot}$ reach higher degrees 
of obscuration during the phases preceding the C-star stage, compared to their $2-3M_{\odot}$ counterparts (see discussion in Sect.~\ref{dustproduction}). The lower the metallicity the 
shorter the redwards extension of the evolutionary tracks, owing to the lower amounts of 
silicon present in the envelope of low Z stars. This behaviour with mass and 
chemical composition is discussed in detail in \citet[][Fig.~8]{flavia15a}.  
The position of 
this class of objects in the various planes in mainly determined by the 
optical depth, which allows us to draw theoretical obscuration sequences, indicated with 
solid lines in Fig.~\ref{fcmd}.

The SED of the stars indicated with magenta triangles confirms the theoretical expectations, as it is reproduced 
by assuming a majority ($70-90\%$) of silicate 
grains, with lower percentages of alumina dust and solid iron and luminosities in the range 
$5\times 10^3-10^4~L_{\odot}$ (see Table 2 for details). We report in Fig.~\ref{flowm} two examples 
of these objects, for which we show the {\itshape Spitzer} IRS data, the photometry available in the literature 
and the best fit model. These two cases exhibit significantly different degrees of obscuration that 
cover almost the whole range of the optical depths derived.

This range covered by the luminosities and the position of these stars in the PL
plane shown in Fig.~\ref{pl} suggests that these sources form a homogeneous group, made up
by objects that descend from low-mass progenitors \citep[in agreement with][]{trabucchi18},
the progeny of $M \leq 1.5~M_{\odot}$ stars, older than $\sim 2$ Gyr.
Indeed stars of mass $2~M_{\odot} \leq M \leq 3~M_{\odot}$ form negligible quantities of 
dust during the M-stage (see discussion in Sect.~\ref{spettro}).


\begin{figure}
\begin{minipage}{0.48\textwidth}
\resizebox{1.\hsize}{!}{\includegraphics{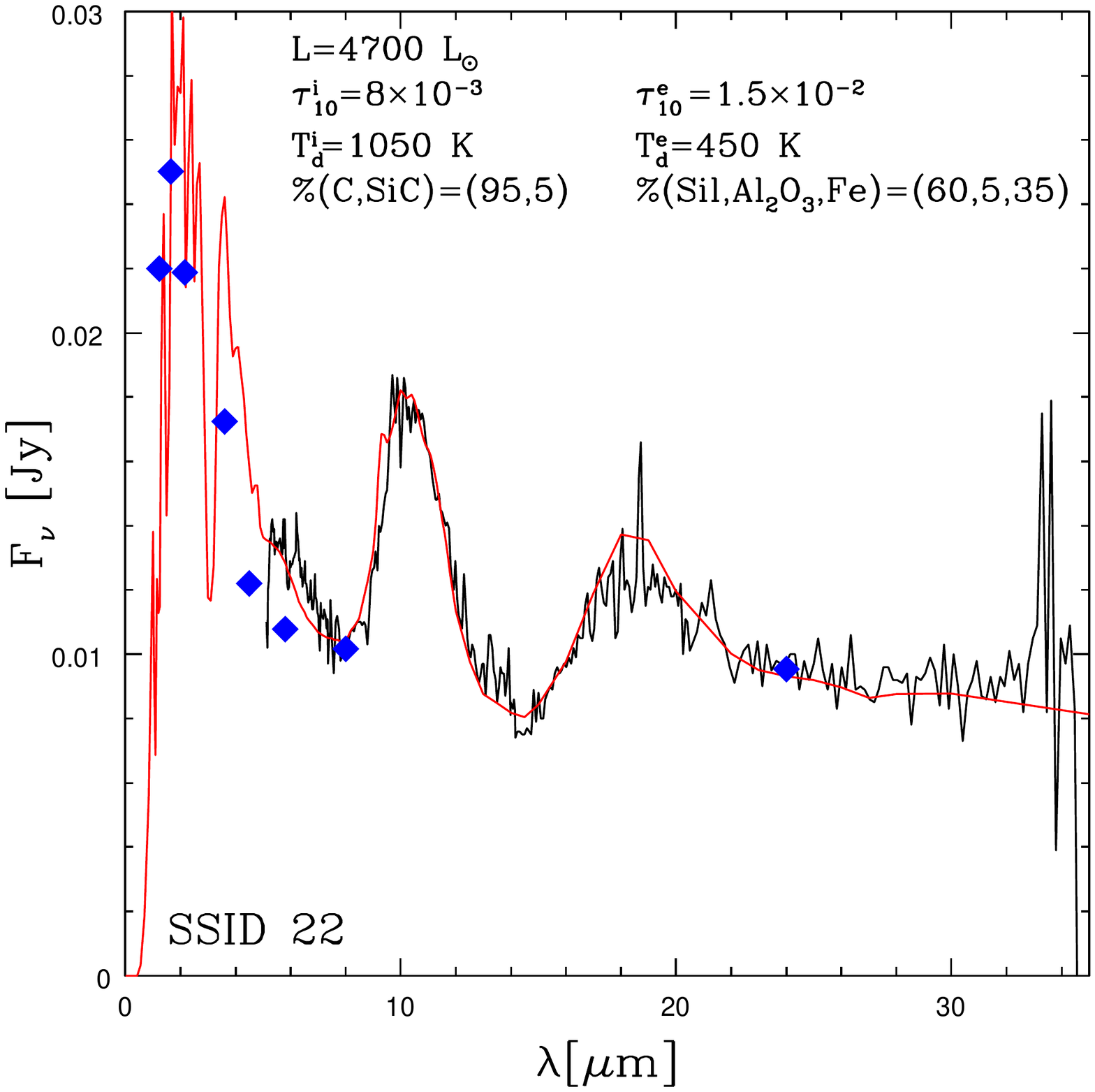}}
\end{minipage}
\vskip-60pt
\caption{Best-fit model for the source SSID22 (same symbols and colours as in 
Fig.~\ref{flowm} are used), interpreted as a low-mass star that has just reached 
the C-star stage (see text for details). The values of the overall luminosity and
of the parameters of the internal and external dust layers are indicated.}
\label{fcstar}
\end{figure}

In summary, we base the characterization of the stars in this sub-sample on the following points: 
i) the luminosities are below $\sim 10^4~L_{\odot}$; ii) dust 
is mainly composed of silicates; iii) the optical depths are in the range 
$0.005 < \tau_{10} < 0.1$.

The range of optical depths and dust composition required to reproduce the SED are in 
agreement with the low-mass models discussed in \citet{flavia15a}. The fraction of silicates
is found to increase across the obscuration sequence, ranging from $\sim 70\%$ to
$\sim 90\%$.
Alumina dust is more stable than silicates, thus it forms at larger rates, in a more 
internal region of the circumstellar envelope, at temperatures of the
order of $\sim 1400$ K; silicates form in a more external zone, at temperatures
$T \sim 1100$ K \citep{flavia14a}. The increase in wind density triggers a 
higher formation rate of both species; however, this has a larger effect on the amount of 
silicates formed, as alumina dust is so stable that it forms in quantities close to 
saturation, thus being less sensitive to variations in the thermodynamics of the wind.

The reddest objects of this group among the obscuration sequence in the ([F770W]-[F2550W], 
[F770W]) plane, with [F770W]-[F2550W]$>2$ and $\tau_{10} \sim 0.1$, are evolving through 
the final phases as M stars, before becoming carbon stars. This hypothesis was proposed by 
\citet{flavia15a} to explain the O-rich stars studied by \citet{blum06}, that define a 
prominent finger in the colour-magnitude ([8.0]-[24],[24]) diagram.

The present interpretation might be tested by measuring the surface C$/$O ratio in these
stars (e.g. via near-IR spectroscopy), that is expected to increase across the sequence, 
until becoming slightly below unity for the sources exhibiting the largest degree of 
obscuration.

In this context, the sources SSID 22 and SSID 130, indicated with yellow open
diamonds, represent an ideal prosecution of this evolutionary path, because we interpret them
as stars that have only recently reached the C-star stage, with the circumstellar envelope
hosting a more internal, hot, dusty layer, populated by solid carbon particles, and a
cooler zone, where silicate dust formed during previous evolutionary phases is
expanding away from the star. The details of the best fit obtained for SSID 22
are shown in Fig.~\ref{fcstar}. 

The obscuration trend in the ([F770W]-[F2550W], [F770W]) plane 
is approximately horizontal; this behaviour is connected with the evolution of the shape
of the SED of low-mass stars with $\tau_{10}$, shown in the left panel 
of Fig.~\ref{fspet}. With increasing optical depth the height of the silicate feature and 
the overall spectrum in the $\lambda > 10~\mu$m region increase, whereas the [F770W] flux 
keeps approximately constant, owing to the reasons explained above, related to the 
shape of the silicate feature.

Inspection of Fig.~\ref{fcmd} shows that ([F770W]-[F2550W], [F770W]) is the only plane 
where the obscuration sequence of these objects has a significant extension and does not
overlap with different kind of stars. In the ($K_{S}$-[F770W], [F770W]) plane they lay very 
close to the dust-free sources, for the reasons given in Sect.~\ref{nodust}. In the 
([F1000W]-[F1500W], [F1000W]) plane these sources populate a diagonal band, reported in the
bottom, left panel of Fig.~\ref{fcmd}, with the most obscured stars located in the bluer 
and brighter region. This is due to the increasing prominence of the $10~\mu$m feature, 
that rises the [F1000W] flux. The colour range, $\Delta$([F1000W]-[F1500W]) $\sim 0.5$ 
mag, is less extended than in the previous cases, because the increase in the flux in the 
$10~\mu$m spectral regions is accompanied by the rise of the flux in the whole region at 
wavelengths $\lambda > 10~\mu$m (see left panel of Fig.~\ref{fspet}). 

In the ([F1000W]-[F2100W], [F1000W]) diagram the identification of these stars is even harder, 
because the percentage increase in the flux as the optical depth increases are very 
similar in the spectral regions where the two filters are centered, i.e. $10$ and
$21~\mu$m.

\begin{figure*}
\begin{minipage}{0.48\textwidth}
\resizebox{1.\hsize}{!}{\includegraphics{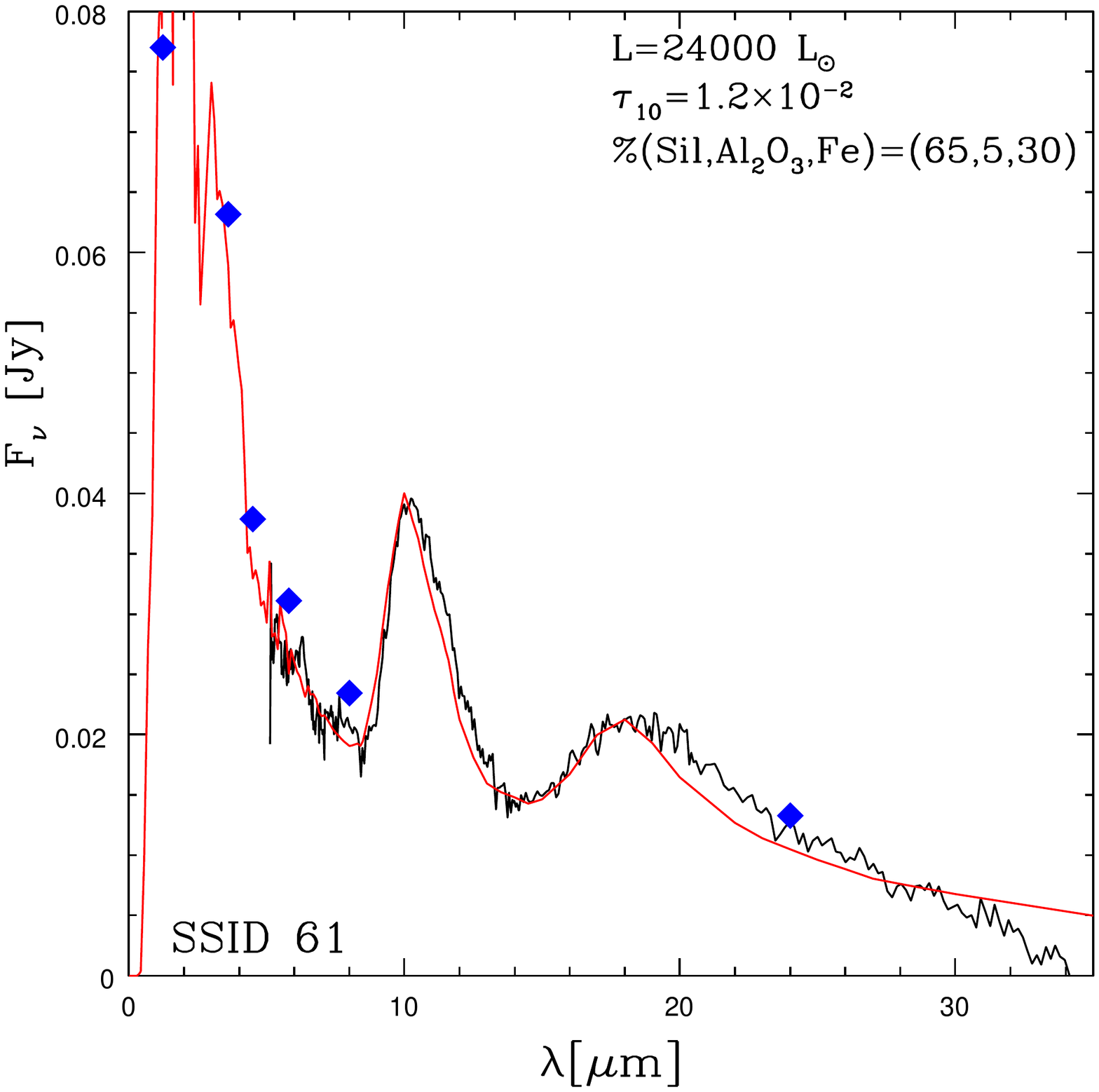}}
\end{minipage}
\begin{minipage}{0.48\textwidth}
\resizebox{1.\hsize}{!}{\includegraphics{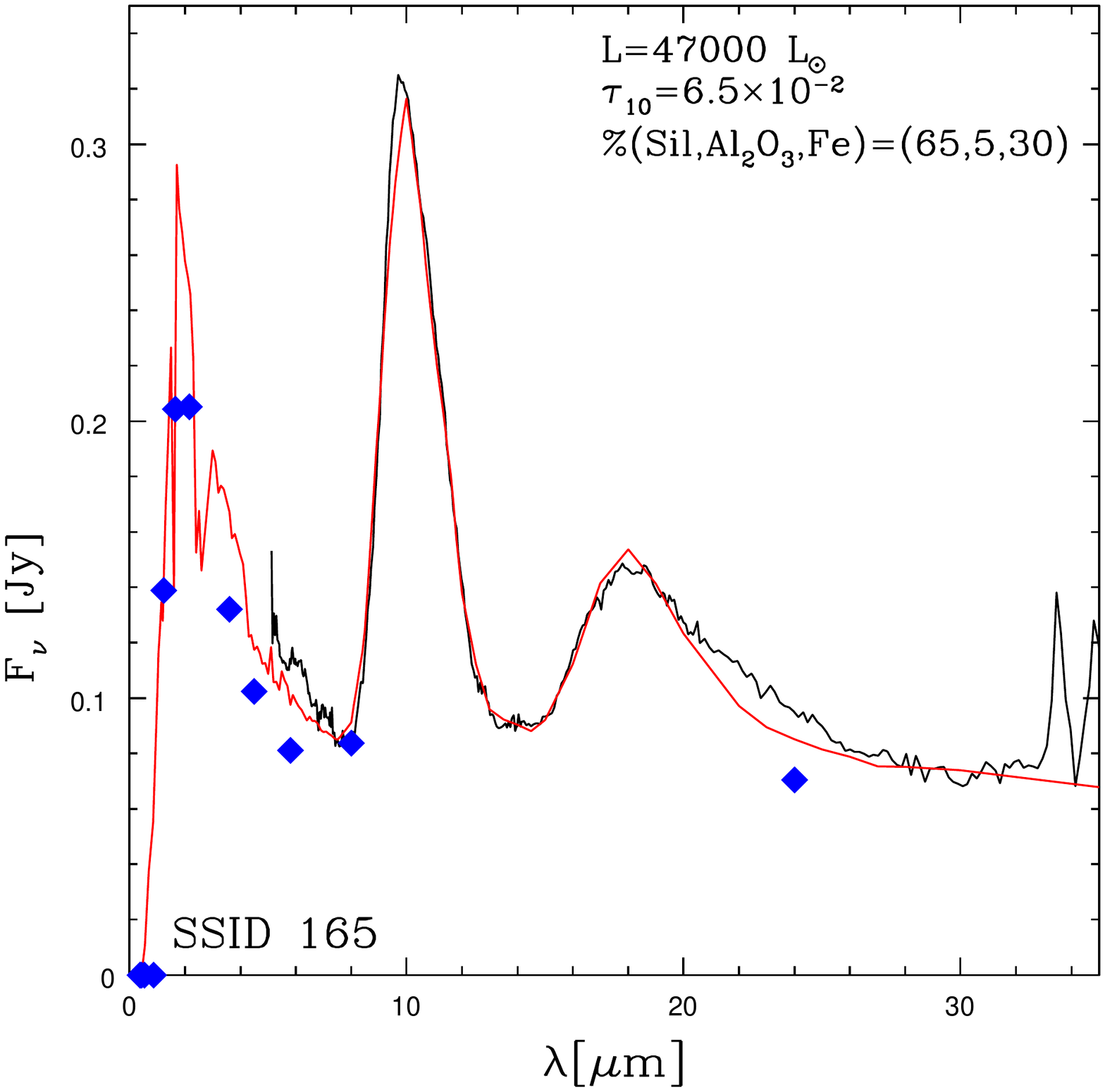}}
\end{minipage}
\vskip-80pt
\begin{minipage}{0.48\textwidth}
\resizebox{1.\hsize}{!}{\includegraphics{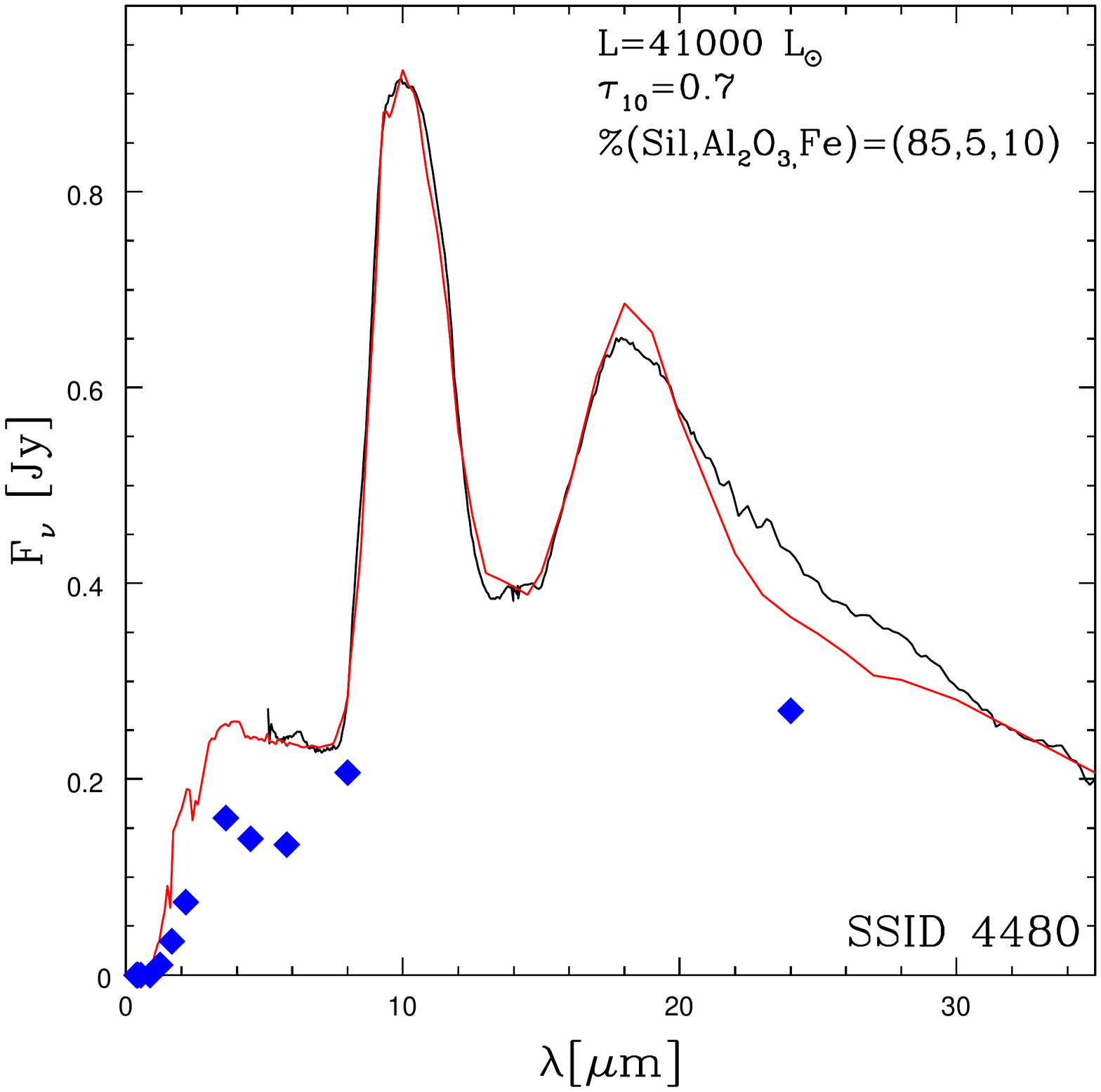}}
\end{minipage}
\begin{minipage}{0.48\textwidth}
\resizebox{1.\hsize}{!}{\includegraphics{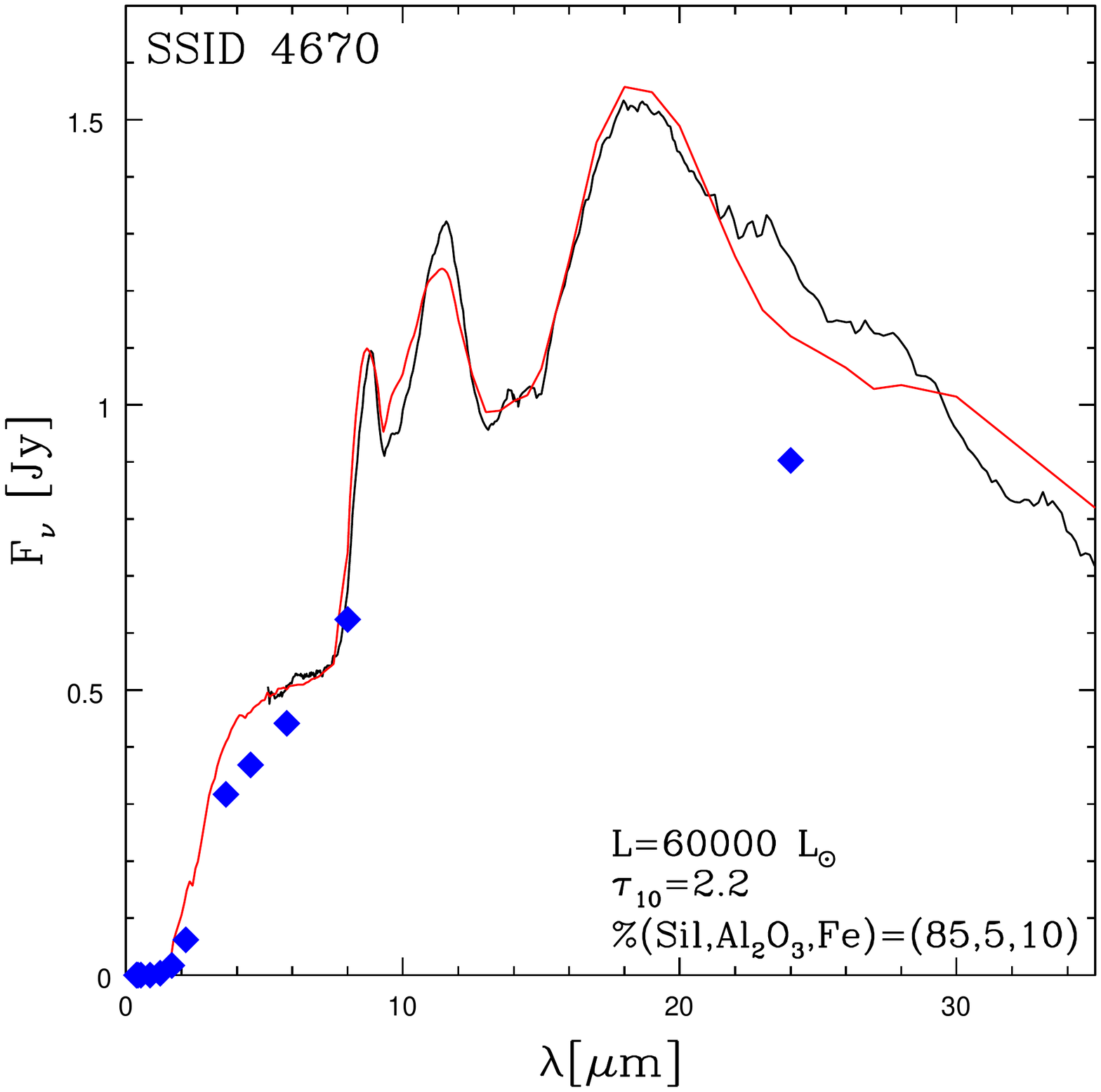}}
\end{minipage}
\vskip-60pt
\caption{The {\itshape Spitzer} IRS spectrum (black line) of 4 stars, selected among those interpreted
as massive AGBs, currently experiencing HBB, taken along the obscuration sequence 
indicated with a dashed line in Fig.~\ref{fcmd}. The photometry and the best-fit
are shown, with the parameters adopted.}
\label{fhbb}
\end{figure*}

\subsection{Stars undergoing hot bottom burning}
\label{hbbstars}
We now turn the attention to the stars indicated with green squares in Fig.~\ref{fcmd}. A 
few examples of the SED of these objects are shown in Fig.~\ref{fhbb}, where we present the 
observations and our corresponding theoretical SED.

Unlike low-mass stars, massive AGBs span a wide range of luminosities, extending by a factor
$\sim5$ (see left panel of Fig.~\ref{fmod}). The position of each model in the various planes is  
determined by the values of the luminosity and optical depth. Therefore the tracks of stars 
of different masses are
practically overlapped, the only difference being that the higher the mass of the progenitor the larger the
maximum luminosity and degree of obscuration reached (see left panel of Fig.~\ref{fmod}), and the wider the excursion
of the evolutionary track in the planes. The dashed lines shown in Fig.~\ref{fcmd} connect points characterized 
by different optical depths and luminosities, therefore can be considered as obscuration sequences of massive AGBs.

The extension of the obscuration sequences is at first order independent of metallicity. 
On the other hand, on the statistical side, we must consider that for lower metallicities 
the range of masses that reach the largest degrees of obscuration is narrower and, more 
important, the largest optical depths are experienced only during a small fraction of the 
AGB life \citep[see Fig.~2 in][for the metallicity effects on the obscuration 
of massive AGBs]{flavia15b}.

The luminosities deduced from the analysis of the SED are in the range $2-8\times 10^4~L_{\odot}$
and the optical depths span the interval $\tau_{10} \sim 0.005-3$. 
The SED of these sources can be nicely interpreted by assuming a "standard" dust mineralogy 
for obscured M stars: a dominant contribution from silicates, with percentages in the range 
$60-100 \%$, completed by smaller fractions of alumina dust and solid iron. The higher the 
required fraction of silicate the larger the optical depth, in 
agreement with the arguments presented earlier in this section. 

These results, compared with the theoretical models discussed in Sect.~\ref{models}
(see left panel of Fig.~\ref{fmod} and the sequence of SEDs in the right panel of 
Fig.~\ref{fspet}), indicate that these stars descend from progenitors of mass above 
$3~M{\odot}$, formed not earlier than $\sim 250-300$ Myr ago. This is also confirmed by 
their position in Fig.~\ref{pl}: their PL relation is in agreement with \citet{trabucchi18} 
models for massive AGB stars ($> 3 M_{\odot}$).

The significant amount of dust present
in the circumstellar envelope rules out the possibility that these stars are metal poor
and suggests that their metallicity is $Z \sim 4-8\times 10^{-3}$; this hypothesis is 
supported by the mass-metallicity relation of the LMC, according to which 
star formation in the last 1 Gyr is dominated by a $Z=8\times 10^{-3}$ stellar
population \citep{harris09}. 

\subsubsection{Massive AGBs in the colour-magnitude planes}
The afore mentioned obscuration sequence defined by massive AGB stars can be used 
to attempt their characterization. Generally speaking, 
we find that the optical depths increase across the 
sequences in the different planes: the sources indicated with open and full green squares are
characterized by $5\times 10^{-3} < \tau_{10} < 0.1$, whereas dotted and crossed green squares correspond
to $\tau_{10} > 0.5$. 

In the ([F770W]-[F2550W], [F770W]) plane (see top, left panel of Fig.~\ref{fcmd}) the 
region populated by these objects extends over $\sim 2$ mag in [F770W]-[F2550W] and 
almost 4 mag in [F770W]. Unlike the lower mass counterparts, discussed in Sect.~\ref{lowmass}, 
the luminosity is generally correlated with $\tau_{10}$, the brighter stars being on the average 
more obscured. The stars indicated with green, dotted squares are an exception to this rule,
as they are fainter than their counterparts with similar degree of obscuration, indicated
with green, crossed squares. We will discuss these sources later in this section.

Both the [F770W] and [F2550W] fluxes rise with increasing $\tau_{10}$, as 
shown in the right panel of Fig.~\ref{fspet} and in the examples in Fig.~\ref{fhbb}. 
On the other hand, it is clear from Fig.~\ref{fspet} that the increase of the flux in 
the spectral region around $25.5~\mu$m is percentually larger when compared to the 
$\lambda \sim 7.7~\mu$m zone, thus making [F770W]-[F2550W] to increase with $\tau_{10}$, 
provoking a rightwards trend of the obscuration sequence.

The stars discussed here trace a diagonal pattern in the ($K_{S}$-[F770W], [F770W]) plane.
With the excursion in the [F770W] magnitude discussed before, 
$K_{S}$-[F770W] spans a range of almost 6 mag, due to the gradual shift of the whole SED
towards mid-IR wavelengths and the decrease in the near-IR flux, particularly relevant for 
$\tau_{10} > 0.1$. This allows for a higher sensitivity to the degree of obscuration, 
although the measurement of the near-IR flux could be critical for the most obscured stars.
Indeed one out of the 3 brightest stars are not reported in this plane, owing to
the lack of the $K_{S}$ flux.

Regarding the ([F1000W]-[F1500W], [F1000W]) diagram, shown in the bottom, left panel of 
Fig.~\ref{fcmd}, the path traced by the obscuration pattern presents a 
turning point: the initial trend towards the blue, down to [F1000W]-[F1500W] $\sim -0.2$, 
is followed by a redwards excursion. The first part is due to the appearance 
of the feature at $9.7~\mu$m, that rises the $10~\mu$m flux and diminishes the emission in the 
wavelength region around $15~\mu$m (see Fig.~\ref{fspet}). When $\tau_{10}$
exceeds $\sim 0.1$ this trend is reversed, because the whole mid-IR flux
is lifted. The most obscured stars distribute approximately horizontally on this plane, 
because for $\tau_{10} > 2$ the silicate feature turns into absorption, thus no further 
increase in the $10~\mu$m flux occurs. The stars with the highest degree of
obscuration that correspond to the brightest and youngest objects discussed previously,
are located in the right, upper region of the plane.

In the ([F1000W]-[F2100W],[F1000W]) diagram the sequence of the stars discussed here follows 
a slightly different behaviour compared to the ([F1000W]-[F1500W],[F1000W]) plane. The obscuration 
pattern extends towards higher [F1000W] fluxes; however, no clear turning point is found,
because the rise of the SED in the $10$ and $21~\mu$m spectral regions occur with 
similar percentages. The stars with $\tau_{10} > 0.5$ populate the brightest region of 
this plane, at [F1000W]-[F2100W] $> 1$.

\subsubsection{Stars undergoing soft HBB}
The stars indicated with open squares harbour little amounts of dust in their 
circumstellar envelope. The optical depths are a few
$10^{-3}$. According to our interpretation, they have just started 
the HBB activity. We expect that only a modest depletion of the overall surface carbon has 
occurred and that the $^{12}$C$/^{13}$C ratio has dropped to values close to the 
equilibrium abundances, of the order of $\sim 4$. No meaningful depletion of the surface 
oxygen is expected.

Following the theoretical obscuration pattern, we analyze the sources indicated with full
squares in Fig.~\ref{fcmd}, that have
$0.01 < \tau_{10} < 0.1$ and luminosities covering the interval $2-4.5\times 10^{4}L_{\odot}$. 
The sources SSID 61 and SSID 165, reported in the top panels of Fig.~\ref{fhbb}, belong 
to this group. The luminosities and $\tau_{10}$'s given above suggest that these objects 
are the progeny of $4-5~M_{\odot}$ stars formed $100-300$ Myr ago, currently experiencing HBB. 
As shown in the middle panel of Fig.~\ref{fmod}, the surface carbon should be $\sim 10-20$ 
timessmaller than in their less obscured counterparts, the surface nitrogen enhanced 
by one order of magnitude and the surface $^{12}$C$/^{13}$C ratio should be very 
close to the equilibrium value, i.e. $3-4$. Furthermore, these stars are expected to be 
experiencing the Cameron-Fowler (1971) mechanism, thus they should be enriched in lithium.

\subsubsection{Highly obscured O-rich AGBs}
The objects indicated with crossed and dotted green squares in Fig.~\ref{fcmd}
are producing dust at a high rate, as confirmed by the large optical depths,
$\tau_{10} > 0.5$.  

In the ([F770W]-[F2550W], [F770W]) and ($K_S$-[F770W], [F770W]) planes
the obscuration patterns are most easily distinguished. The stars with [F770W]$>5.5$ 
(all the crossed and dotted, green squares, but the 3 brightest objects) have luminosities below 
$\sim 5\times 10^4~L_{\odot}$; an example is 
the source SSID 4480, reported in the bottom, left panel of Fig.~\ref{fhbb}. 
Based on the results shown in the left panel of Fig.~\ref{fmod}, and on the discussion 
in Sect.~\ref{properties}, 
we deduce that they descend from progenitors of mass in the range $3.5-5~M_{\odot}$, 
formed between 100 Myr and 300 Myr ago. The optical depths, 
$0.5 < \tau_{10} < 1.5$, are consistent with this conclusion. 

The 3 brightest sources with [F770W]$<5.5$ are characterized by a high degree of obscuration, 
with $\tau_{10}\sim 2-3$, and luminosities above $5\times 10^4~L_{\odot}$, that, based on 
the results shown in the left panel of 
Fig.~\ref{fmod}, indicate progenitors of mass above $\sim 5~M_{\odot}$, younger than
$\sim 100$ Myr. 

A word of caution regarding dust formation
modelling in the winds of these peculiar objects is needed here. While AGB models
reproduce both the luminosities and the periods of these sources, the optical depths
required, in the range $2 < \tau_{10} < 3$ are higher 
than the largest theoretical values predicted, of the order of $\tau_{10} \sim 1$. This 
outlines some tension between models and observations. Recent radiation-hydrodynamic (RHD) models of winds of M-type AGB 
stars \citep[see, e.g.,][]{hofner16,hofnerolof18} indicate that dust formation will indeed 
start deeper inside the atmosphere (at smaller condensation radius), but this result was 
obtained at solar metallicity (not typical LMC metallicity). However, the difference in 
condensation radii in the simple stationary outflows of the stellar evolution models and 
the more sophisticated RHD models, is likely a result of the different levels of detail of 
the two types of models. First, the RHD models are known to produce different results due 
to the dynamics (pulsation) and time dependence. Second, frequency-dependent radiative 
transfer (RT) changes the energy balance and temperature structure compared to the 
atmospheric structures obtained in the stellar evolution models. That is, the simplistic 
atmosphere models we employ probably do not predict correct condensation radii. 

\subsubsection{A class of massive AGBs in the very late AGB phases?}
A second issue of the present understanding is the interpretation of the stars
indicated with dotted, green squares in Fig.~\ref{fcmd}. These sources have luminosities 
of the order of $2\times 10^4~L_{\odot}$, significantly smaller than those of their brighter 
counterparts, indicated with crossed, green squares. 
These luminosities are compatible both with those of stars evolving through the initial TPs, or at the 
end of their AGB life, after HBB was turned-off. Their periods (we refer to the OGLE period for SSID 4007
and those from GS18 for the other sources in this group) are close or above 1000 d. 
Because in the first case we should expect periods below $\sim 500$d (see Fig.~\ref{pl}), we believe more
plausible that these stars are experiencing the latest AGB phases.

The main drawback of this interpretation is that their degree of
obscuration is also expected to be low (see the bottom, left panel of Fig.~\ref{fcmd}),
whereas their optical depths are $\tau_{10}\geqslant1$. A possible
solution is that during the very final AGB phases the formation of a disc favours the 
accumulation of the dust produced during earlier phases, such that the overall dust density is
significantly higher than predicted by the simplified description of the wind used in the
present analysis. We leave this problem open.

The surface chemical composition of the stars indicated with crossed and dotted, green squares is expected to 
show the imprinting of proton-capture processing, in analogy with the less obscured 
counterparts, indicated with open and full, green squares. These stars are also expected to have started the 
Ne-Na nucleosynthesis, with a sodium enrichment by a factor of $\sim 5$ \citep{ventura13}.
If this small sample includes stars of lower metallicity, of the order
of $Z=4\times 10^{-3}$, then we would observe the results of oxygen burning and of Mg-Al 
nucleosynthesis, with a surface Al enhancement by a factor $5-10$ and oxygen depletion
by a factor $\sim 2$ \citep{ventura16}.

\begin{figure*}
\begin{minipage}{0.48\textwidth}
\resizebox{1.\hsize}{!}{\includegraphics{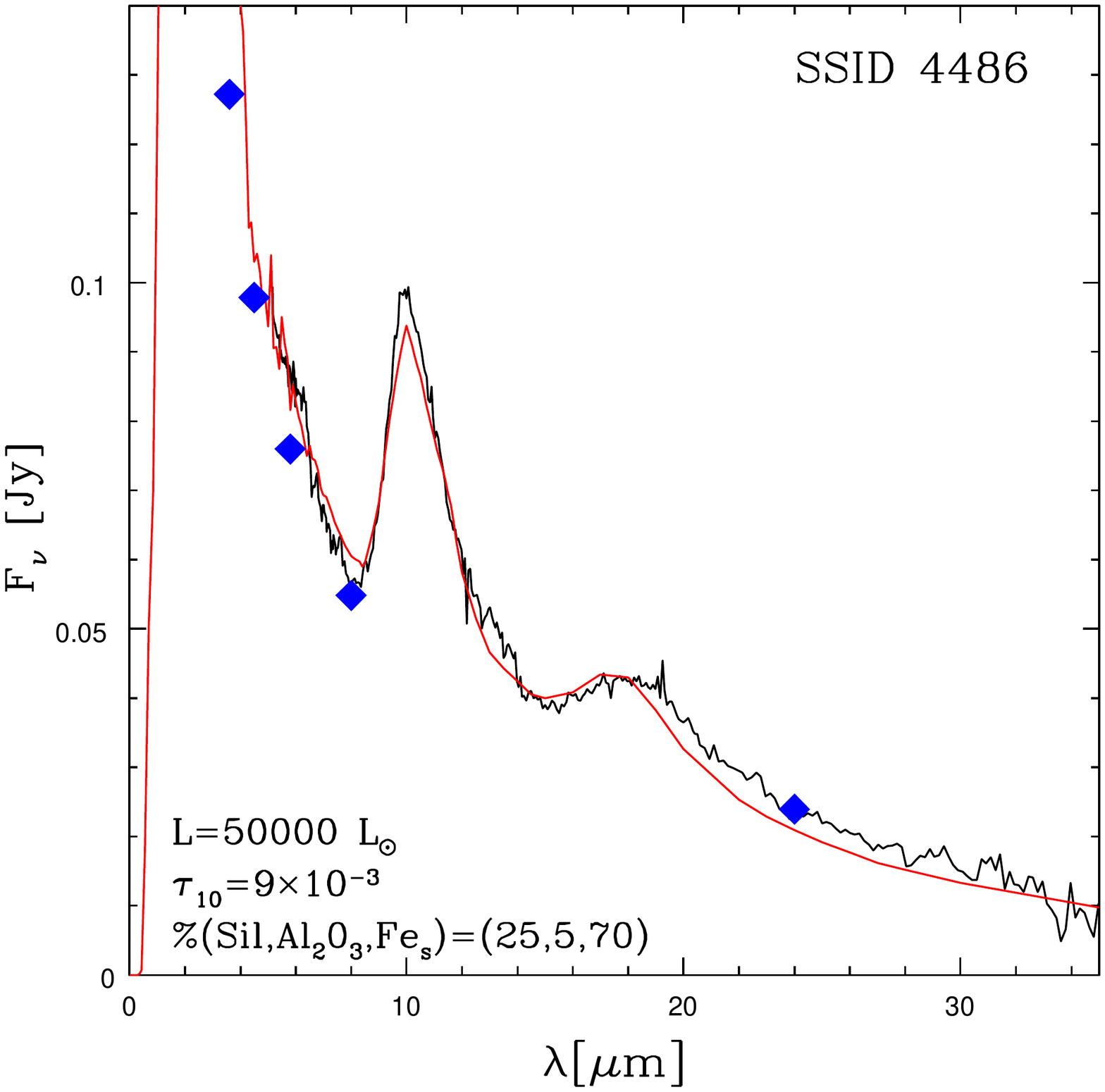}}
\end{minipage}
\begin{minipage}{0.48\textwidth}
\resizebox{1.\hsize}{!}{\includegraphics{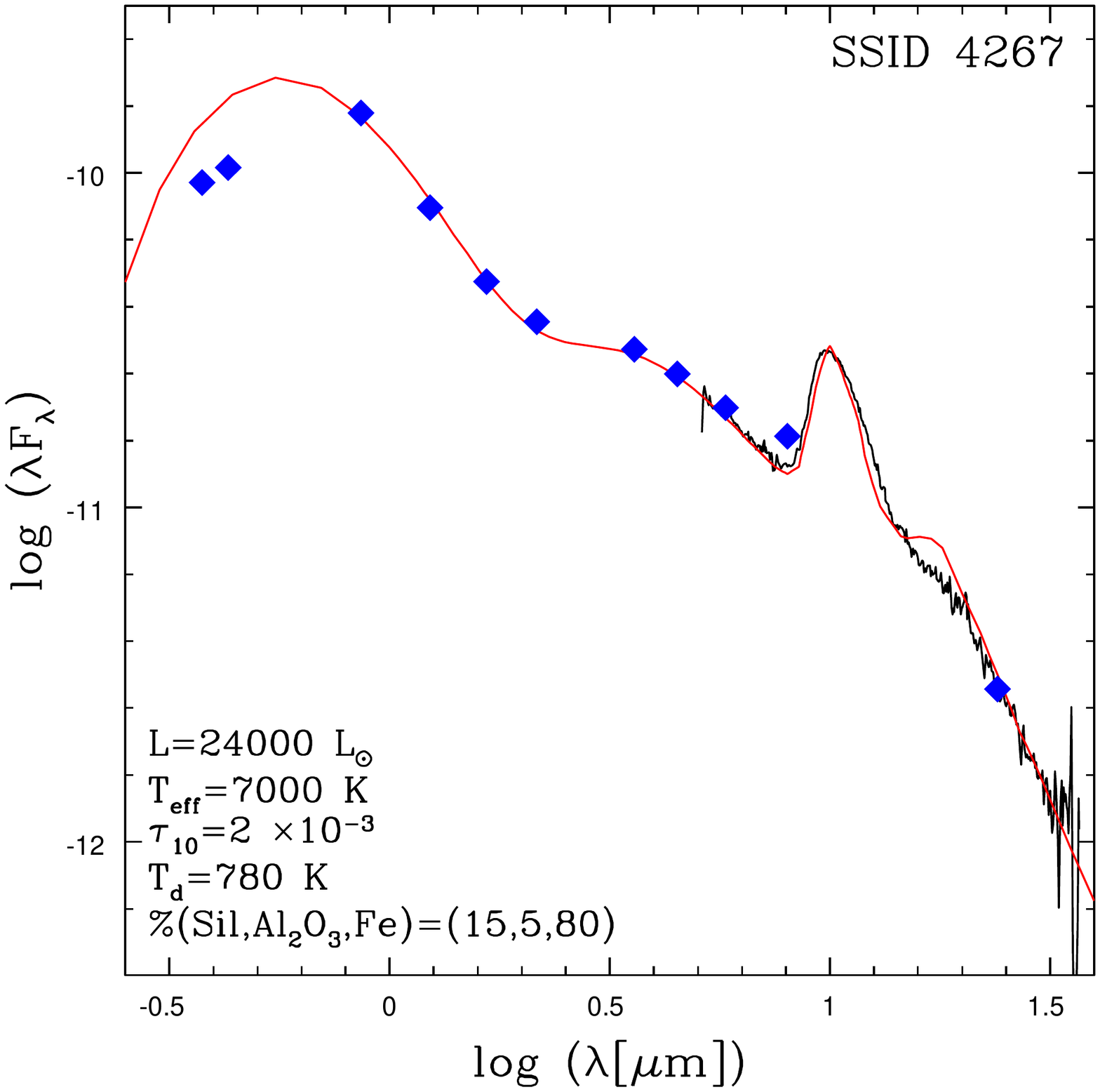}}
\end{minipage}
\vskip-60pt
\caption{IRS, photometry results and best-fit models for the sources SSID 4486 (left panel)
and SSID 4267 (right), interpreted as stars surrounded by dust, whose composition is dominated by
iron grains, evolving through the AGB and the post-AGB phase, respectively. For SSID 4267
we report the flux in the $\log (\lambda F_{\lambda})$ scale, to better understand why this
object was misclassified as an AGB star (see text for a discussion on this point).}
\label{firon}
\end{figure*}

\subsection{Iron-dust stars}
\label{festars}
\subsubsection{Massive AGB stars with iron dust}
The sample studied by \citet{jones12, jones14, jones17} include a paucity of bright objects, indicated
with blue crosses in Fig.~\ref{fcmd}, whose luminosities are above $4.5\times 10^4~L_{\odot}$. 
The large fluxes indicate that these stars are the progeny of $M>4~M_{\odot}$ stars,
currently experiencing HBB. The peculiarity of these stars is in their SED, that exhibits
the main silicate feature at $9.7~\mu$m, and cannot be reproduced by assuming a 
silicate-dominated dust mineralogy. An example of such a SED is shown in the left panel of 
Fig.~\ref{firon}. Unlike the bright stars undergoing HBB, discussed earlier in this section,  
the degree of obscuration is not particularly large, with $\tau_{10} < 0.01$. 

\citet{ester19a} proposed that the stars in this group are the progeny
of metal-poor, $M > 4~M_{\odot}$ stars, formed $\sim100$ Myr ago. The little amount of 
dust present in their surroundings is due to the low metallicity, of the order of 
$Z \sim 10^{-3}$. According to \citet{ester19a} the peculiar SED is related to the strong HBB
experienced, that provoked a significant drop in the surface abundances of oxygen 
and magnesium, and the consequent shortage of water molecules and of magnesium atoms,
both essential ingredients to form silicate particles. The mineralogy of the dust around
these stars is therefore dominated by solid iron, the latter species being not affected
by HBB. Iron grains account for $\sim 80\%$ of the total dust, with smaller
percentages of silicates ($\sim 15\%$) and alumina dust $\sim 5\%$. 

As shown in Fig.~\ref{fcmd}, in the ([F770W]-[F2550W], [F770W]) plane these objects populate 
the region at $1 <$ [F770W]-[F2550W]$<1.5$ and [F770W] $\sim 8$\footnote{The only exception 
is the source SSID 4098, with $ [F770W]-[F2550W] \sim 2.4$, which is 
evolving through a less advanced evolutionary phase, when oxygen in the envelope has not 
been completely burnt, so that $\sim 40\%$ of dust is made up of 
silicates.}. The colours of these sources are not extremely red,
consistently with the low degree of obscuration. The [F770W] fluxes are higher than those 
exhibited by the more metal rich counterparts of similar colours. As discussed in
\citet{ester19a}, this is due to the peculiar shape of their SED, because the
low percentage of silicates prevents the depression of the flux in the spectral region
around $8~\mu$m, thus lifting the [F770W] emission.

In the ($K_{S}$-[F770W], [F770W]) and ([F1000W]-[F1500W], [F1000W]) planes (see Fig.~\ref{fcmd}) 
the iron dust stars cannot be easily identified, as they lie along the obscuration sequence of 
their counterparts of higher metallicity, close to those indicated with full, green squares.

In the ([F1000W]-[F2100W], [F1000W]) plane these sources populate a zone $\sim 0.2$ mag bluer than
the stars with similar optical depths and higher metallicity: this is because the
higher fraction of iron dust affects the SED by decreasing the relative height of the
secondary silicate feature, at $\sim 20~\mu$m, which diminishes the [F2100W] flux,
rendering the [F1000W]-[F2100W] colours bluer.

\subsubsection{Post-AGB stars with iron dust}
\label{fepagb}

In the family of stars surrounded by iron dust we also consider SSID 4267 and 
SSID 6, indicated in Fig.~\ref{fcmd} with open, blue squares filled with crosses. 
The {\itshape Spitzer} IRS and photometric data and the best-fit model for 
SSID 4267 are shown in the right panel of Fig.~\ref{firon}. 

We find that these two objects have effective temperatures in the 
range $T_{\rm eff} \sim 7000-8000$ K, luminosities 
$\sim 2-2.5\times 10^4~L_{\odot}$, and are surrounded by dust, mainly composed
by solid iron. These parameters point to an evolutionary 
nature that is more evolved than that of stars in the AGB phase, suggesting that
they are post-AGBs; this is also consistent with the large IR excess observed
\citep{vanwinckel03, vanaarle11, kamath14, kamath15} and
with the predicted post-AGB luminosities of metal-poor massive AGBs  \citep{marcelo16}. 
SSID 4267 is classified as a post-AGB candidate in the photometric study to search for 
post-AGB stars in the LMC by \citet{vanaarle11}. Additionally, in the Spitzer and TIMMI2 
study by \citet{gielen11a} of post-AGB disc sources, SSID 6 is classified as a post-AGB 
star with atypical dust chemistry, because it shows no strong evidence for the expected 
crystalline features in their spectra. A combination of amorphous silicate dust and 
alumina was required to fit the spectral features in the $5~\mu$m$< \lambda < 40~\mu$m 
region. Since we find that SSID 4267 and SSID 6 are surrounded by iron-dust, we classify 
these two objects as iron-dust post-AGB stars.
 

Iron dust post-AGB stars are characterized by optical depths 
$\tau_{10} \sim 1-5\times 10^{-3}$, smaller than their counterparts evolving 
through the AGB phase. This is because dust formation stops when the 
stars leave the AGB and the dust layer currently observed is expanding 
away from the central star, in a lower density region. The dust 
temperatures required, of the order of $700-900$ K, support this 
interpretation.

In the study by \citet{woods11} these objects have been 
classified as M-type AGB stars. \citet{woods11} identify post-AGB stars 
based on the presence of a distinguished double peak in the SED, separated 
by a clear minimum at $\lambda \sim 2~\mu$m. SSID 4267 and SSID 6 
do not show the expected minimum. Our interpretation for the lack of the 
expected minimum is that the presence of iron dust makes the minimum in the 
SED to vanish almost completely, thus rendering the spectrum similar to a scarcely 
obscured M-type AGB star. Such a modification of the SED is the typical effect of the
presence of significant quantities of a featureless dust species (like solid iron is),
as discussed in the analysis of dual-dust chemistry AGB and post-AGB stars
studied by \citet{ester19b}.

Furthermore, a meaningful difference between the SED of AGB and post-AGB stars 
surrounded by iron dust is the peak in the spectral emission, 
located at shorter wavelenghts in the latter case 
(see Fig.~\ref{firon}, right panel). As a consequence, AGB stars 
have lower [F770W] magnitudes than their post-AGB counterparts,
and their ([F770W]-[F2550W]) colours are bluer. This can be seen in 
the relative position of the two groups of stars in the 
([F770W]-[F2550W], [F770W]) plane. 

($K_{S}$-[F770W], [F770W]) is the plane where iron dust AGB 
and post-AGB stars can be separated more easily. This is 
because the SED of AGB stars generally peaks at wavelengths around 
$\sim 2~\mu$m, that renders their $K_{S}$-[F770W] colours significantly 
bluer than in the post-AGB case.

In the ([F1000W]-[F1500W], [F1000W]) and ([F1000W]-[F2100W], 
[F1000W]) planes iron dust stars populate the same regions.



\begin{figure*}
\begin{minipage}{0.48\textwidth}
\resizebox{1.\hsize}{!}{\includegraphics{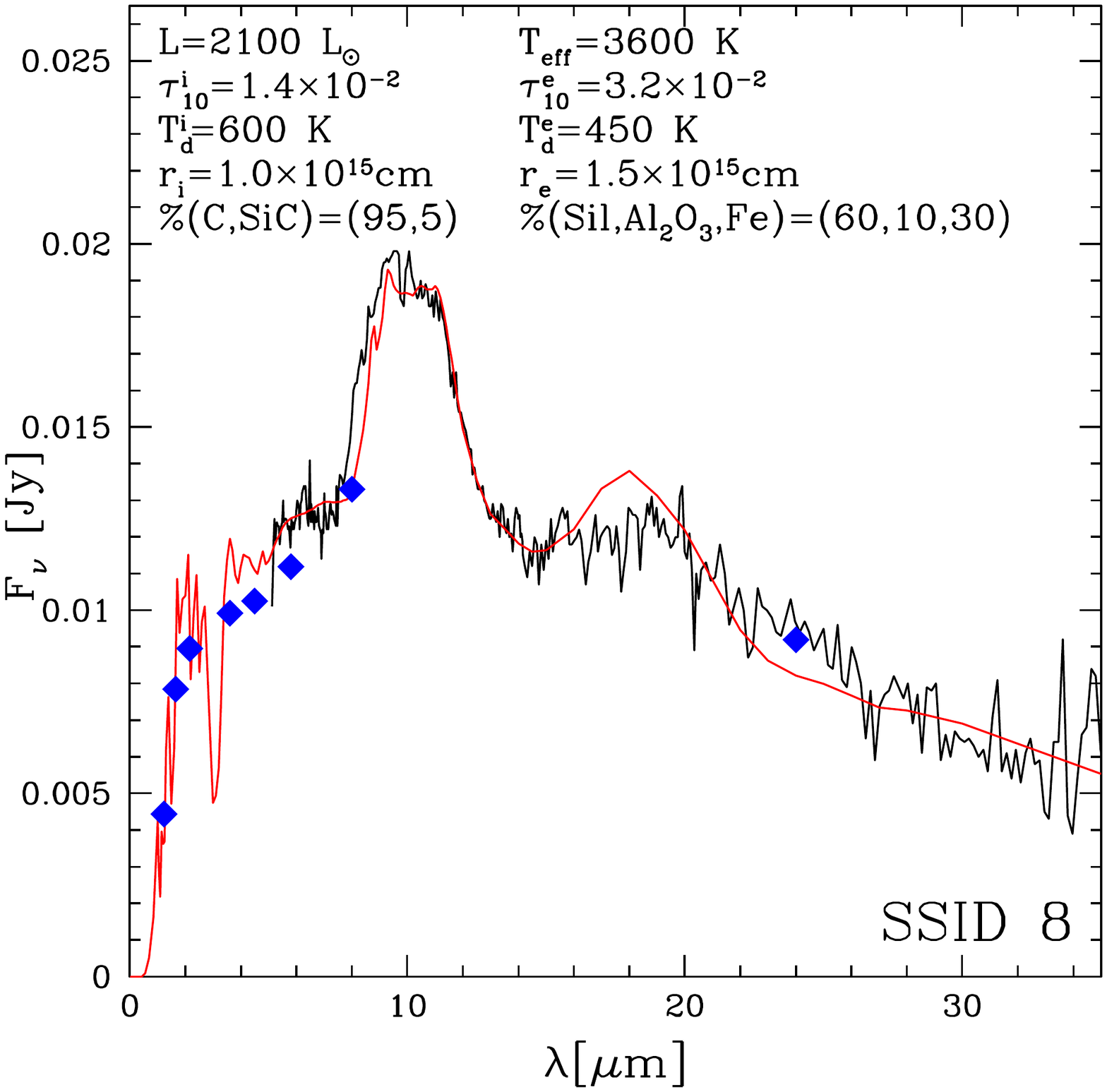}}
\end{minipage}
\begin{minipage}{0.48\textwidth}
\resizebox{1.\hsize}{!}{\includegraphics{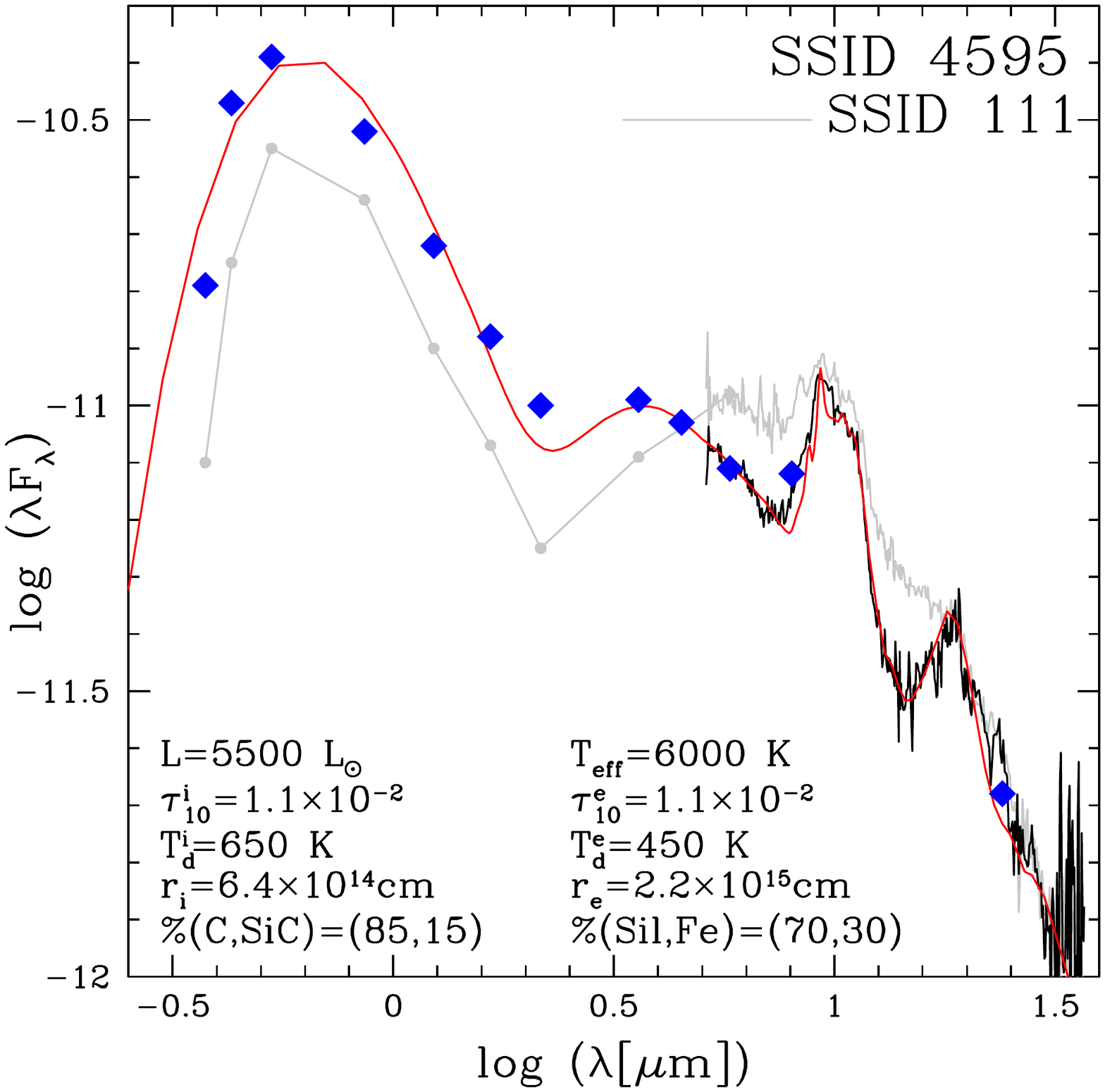}}
\end{minipage}
\vskip-60pt
\caption{IRS, photometric data and best-fit model of two sources interpreted as
stars surrounded by a double-chemistry dust, evolving through a post-thermal pulse 
phase (SSID 8, left) and the post-AGB phase (SSID 4595, right). The parameters of the
two dust layers used to produce the best fit are indicated in the two panels. The
right panel also shows (grey line) the spectrum of a source classified as post-AGB,
following the criterion by \citet{woods11}. It is clear the minimum at $\lambda \sim 2~\mu$m,
usually considered as an indicator of a post-AGB star.}
\label{fdual}
\end{figure*}

\subsection{Dual dust chemistry AGB stars}
\label{dualstars}
The stars indicated with yellow diamonds in Fig.~\ref{fcmd} exhibit a peculiar SED. 
Two examples are shown in Fig.~\ref{fdual}. While the presence of silicate dust is 
indicated by the $9.7~\mu$m and $18.7~\mu$m features, the SED can be hardly accounted 
for by assuming the presence of a single silicate rich dust layer. This lead 
\citet{ester19b} to propose a dual dust chemical composition for these sources. Despite 
according to our interpretation yellow crossed diamonds and full diamonds indicate two 
groups of stars that are evolving through different evolutionary phases, we discuss them 
in the same context, owing to the similarities in the mass and formation epoch of the 
progenitors and in the mineralogy of the dust in their circumstellar envelope.

\subsubsection{Dual-dust post-AGB stars}
The yellow, crossed diamonds (see Fig.~\ref{fcmd}) indicate stars 
(SSID 4054, SSID 4245, SSID 4547, SSID 4595) characterized by a dual-dust chemistry, 
with an effective temperatures above $5000$ K and luminosities in the range
$5\times 10^{3}-10^{4}~L_{\odot}$. These parameters make them compatible with the 
post-AGB phase of low-mass (below $2~M_{\odot}$) progenitors (see Sect.~\ref{post}). 
This interpretation is consistent with the works by \citet{vanaarle11} and 
\citet{kamath15} who carried out a photometric and spectroscopic classification, 
respectively, of post-AGB stars in the LMC. These studies, based on the near-IR excess, 
classify these sources as likely binary post-AGB stars with a 'disc-type' SED. 



We interpret the SED of these objects as composed by an internal dusty region, 
hosting solid carbon particles, and a cooler and more external zone, populated by silicates, 
alumina dust and solid iron. The evolutionary scenario proposed by \citet{ester19b} is 
that these stars became C-stars during their last TP, after evolving as M-stars during the 
whole AGB life. The two dusty layers, currently expanding away from the central star, are 
populated by particles formed before and after the last TP.
This scenario can work only if these sources did not experience
further TPs after becoming carbon stars. According to the discussion in Sect.~\ref{properties} this is indeed a common behaviour of all the stars of mass 
$1.2~M_{\odot} < M < 2~M_{\odot}$\footnote{The lower limit is due to the fact
that stars of initial mass below $\sim 1.2~M_{\odot}$ never become carbon stars.}.
Therefore, we deduce that these objects formed between 1 and 4 Gyr ago. 

\citet{waters98} first proposed that the circumbinary disc may be responsible for mixed 
chemistry: oxygen-rich dust had been ejected by the AGB star in the past and stored in 
the binary disc, while more recent mass-loss was carbon-rich. This interpretation has since 
been supported by several studies including \citet{gielen11b,matsuura14}.  

\citet{gielen11a}, based on the fit of the spectral region $5~\mu$m$< \lambda < 40~\mu$m, 
concluded that the dust mineralogy for SSID 4595 and SSID 4547 is mainly 
composed of amorphous and crystalline silicates. However, we confirm that for the 
four dual-dust post-AGBs in this study, a significant fraction of carbon dust is required to allow
the fit of the overall SED, especially in the $1~\mu$m$< \lambda < 8~\mu$m region. This
is consistent with the detection of a weak PAH-like emission feature at $6.3~\mu$m in the
SED of SSID 4547 \citep{ester19b}. 

We note that PNe with a similar dual chemistry (i.e. PAHs + amorphous silicates + 
crystalline silicates) are rare but a few of them have been observed towards the Galactic Bulge 
\citep{perea09,gorny10}\footnote{Curiously, the PAHs in these few peculiar PNe are 
very similar to those in SSID 4547, where the PAHs at $6.3$ (and $11.3~\mu$m) are much 
stronger than the other two ($7.7$ and $8.6~\mu$m, almost absent).}. As a confirmation of this, 
SSID 4547 was already observed and classified in \citet{matsuura14} as an oxygen-rich 
post-AGB star with PAH features in its spectrum, raising the possibility that
this object is characterized by a dual-dust chemistry.

\subsubsection{A class of faint AGBs with dual-dust}
Full, yellow diamonds in Fig.~\ref{fcmd} point faint stars with luminosities in the range 
$2-4.5 \times 10^{3}~L_{\odot}$. Their effective 
temperatures indicate that they are AGB sources. It is not expected that such low-luminosity 
AGBs form dust in meaningful quantities (see Fig.~\ref{fmod}). 

The hypothesis that an unusually strong mass loss episode could have temporarily enhanced 
dust formation is unlikely, for the following reasons: i) the dust temperatures are cool, 
thus we are observing stars that are not producing dust in the present time; ii) the SED 
can be reproduced only if some carbon dust is added to silicates, alumina dust and solid iron. 

In analogy with the post-AGB stars, \citet{ester19b} 
reproduced the observed SED claiming the presence of an 
internal dust zone, populated by carbonaceous particles and an outer layer, with a dust
composition dominated by silicates. According to \citet{ester19b} these stars are 
evolving through a post-TP phase, during which the CNO nuclear activity is
temporarily extinguished (which accounts for the low luminosities found), and no dust is 
being formed. The two dusty layers host dust grains
formed during the phases immediately before the TP, when solid carbon grains were formed,
and during the previous inter-pulse phase, before becoming carbon stars.

\subsubsection{Dual-dust stars in the colour-magnitude planes}
Among the different observational planes shown in Fig.~\ref{fcmd}, the diagram where 
dual-dust chemistry objects can be identified most easily is the ($K_{S}$-[F770W], [F770W]) plane.
The presence of carbon dust makes the [F770W] flux higher and the $K_{S}$-[F770W] colour bluer
than in stars of similar optical depth. Therefore, the colours of these stars in this 
plane are comparable with those of the largely obscured stars discussed in Sect.~\ref{hbbstars}, that are much brighter. As shown in the top, right panel of Fig.~\ref{fcmd},
the stars with dual dust chemistry lie in the portion of the ($K_{S}$-[F770W],[F770W]) plane at 
$K_{S}$-[F770W] $\sim 2.5-3.5$, tracing an almost vertical sequence, below the obscuration 
pattern defined by the higher mass counterparts (indicated with green squares). In this 
$K_{S}$-[F770W] colour range is located also SSID 130, that, on the other hand, is 
characterized by a dual dust chemistry as well (see Sect.~\ref{lowmass}).

In the ([F770W]-[F2550W], [F770W]) plane they lie slightly above the stars with similar degree
of obscuration, again for the peculiar shape of the SED in the $\lambda \sim 8~\mu$m spectral
region; however, their identification is more tricky than in the ($K_{S}$-[F770W],[F770W]) plane.

In the ([F1000W]-[F1500W],[F1000W]) and ([F1000W]-[F2100W],[F1000W]) planes the dual 
chemistry stars populate the same region where we find the sources with low obscuration.

\begin{figure*}
\begin{minipage}{0.48\textwidth}
\resizebox{1.\hsize}{!}{\includegraphics{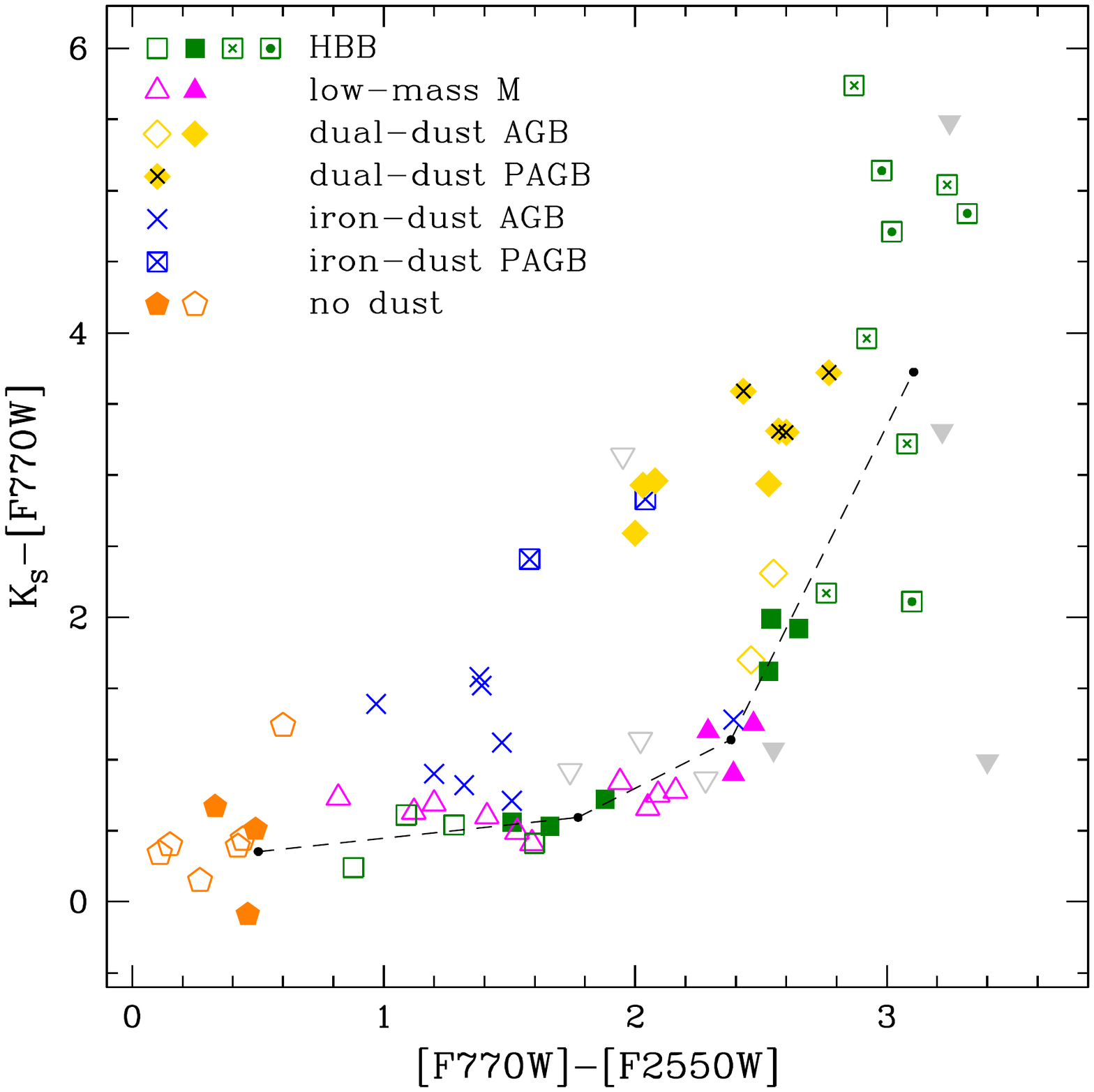}}
\end{minipage}
\begin{minipage}{0.48\textwidth}
\resizebox{1.\hsize}{!}{\includegraphics{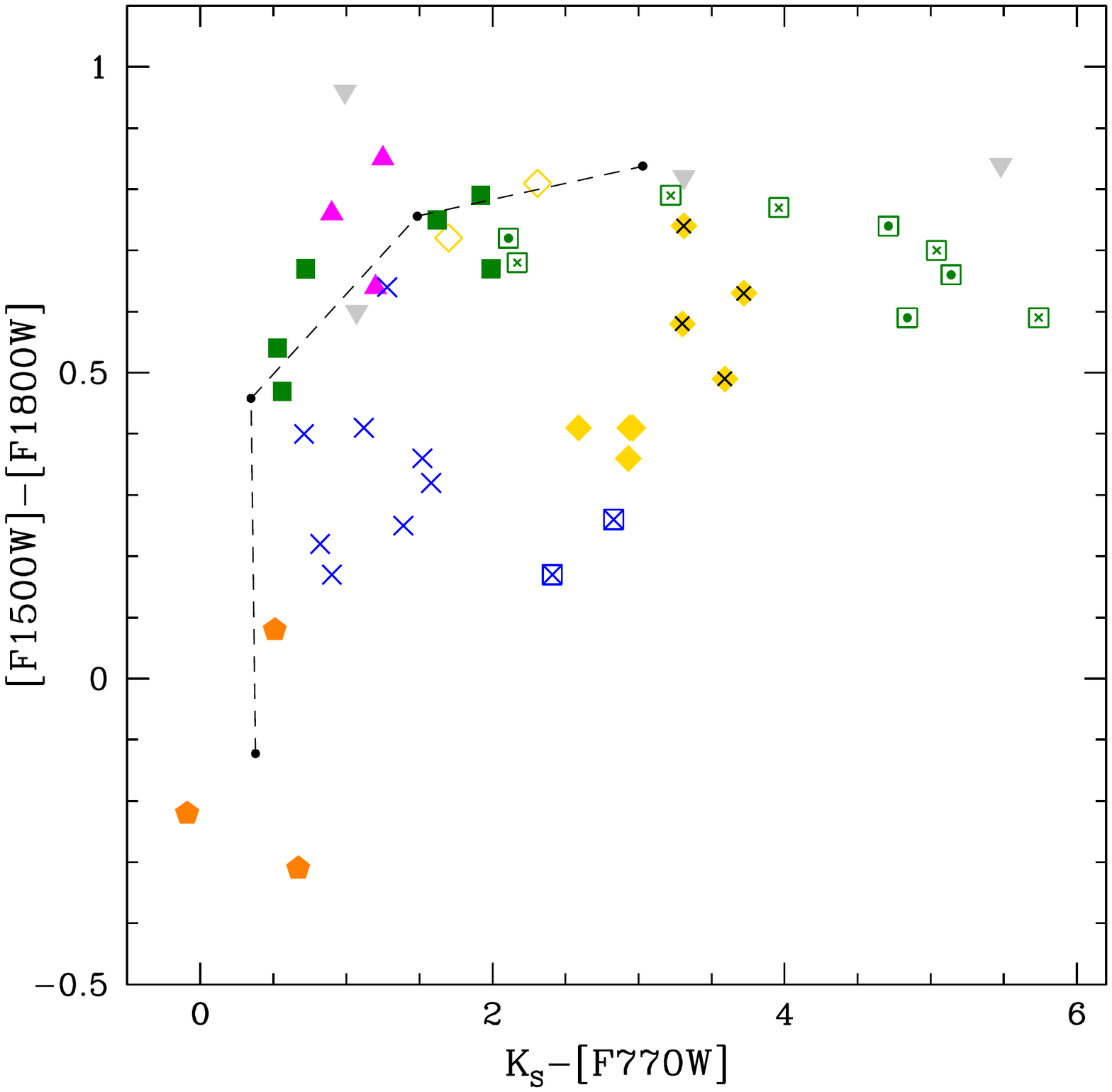}}
\end{minipage}
\vskip-80pt
\begin{minipage}{0.48\textwidth}
\resizebox{1.\hsize}{!}{\includegraphics{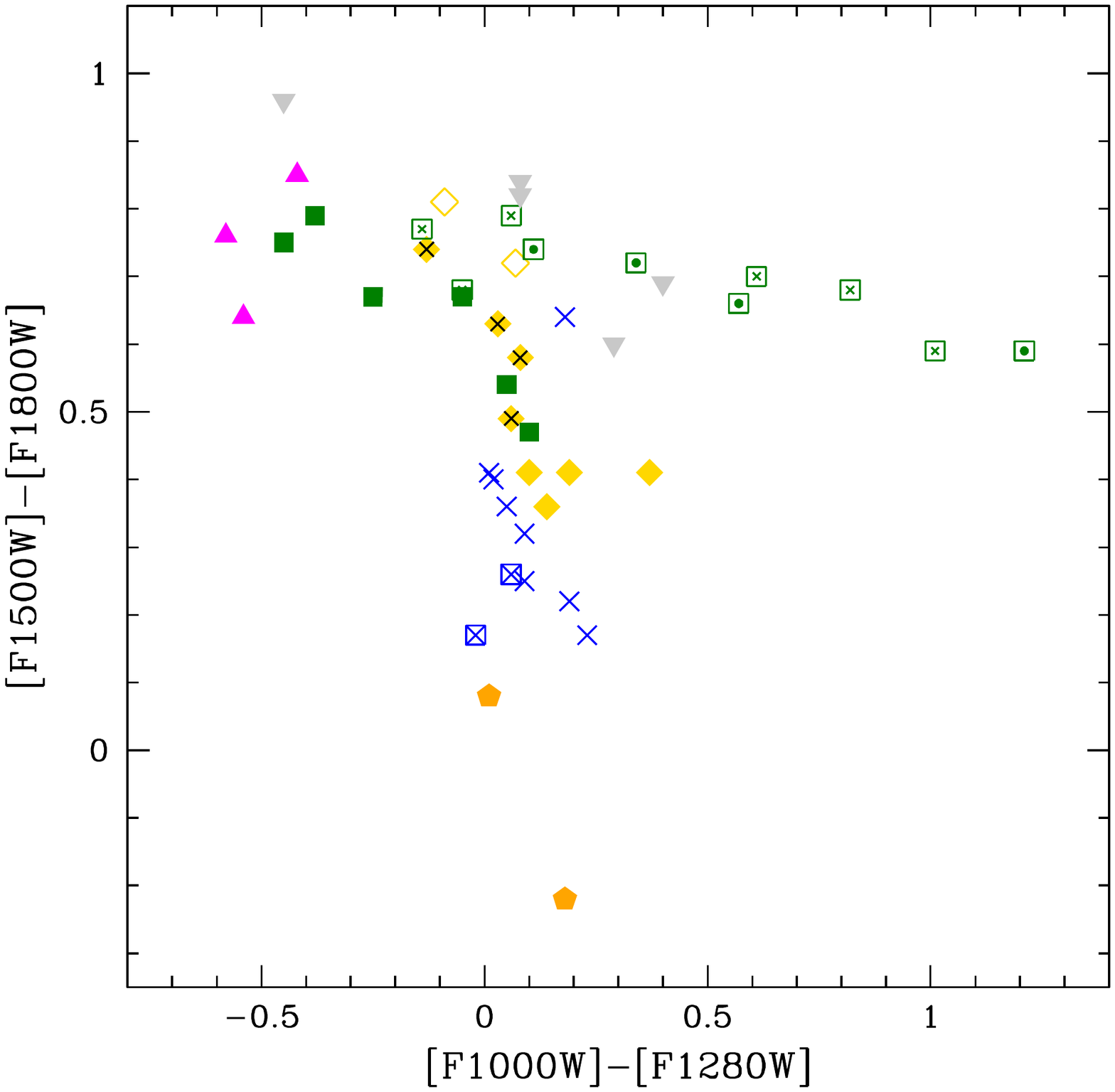}}
\end{minipage}
\begin{minipage}{0.48\textwidth}
\resizebox{1.\hsize}{!}{\includegraphics{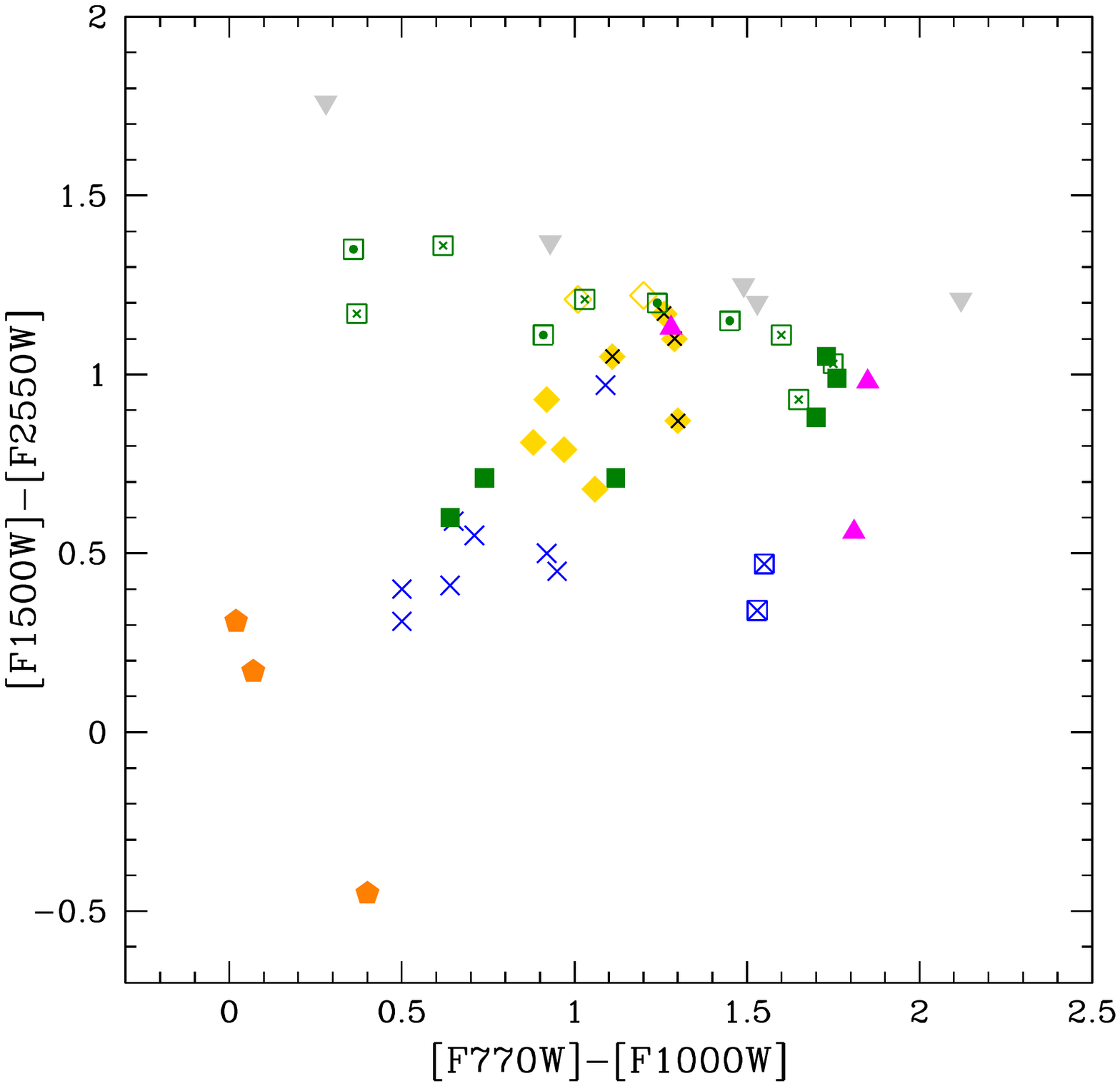}}
\end{minipage}
\vskip-60pt
\caption{The evolved,  M-type stars in the sample published by \citet{jones12, jones14, jones17} in the
various colour-colour planes, obtained by using the IR filters of the {\itshape JWST}.}
\label{fccd}
\end{figure*}

\section{M-type evolved stars in the colour-colour diagrams}
The distribution of the several M stars classes discussed in the previous section
in the colour-colour diagrams can be used to identify the different obscuration
sequences and to select those planes where the various groups can be distinguished more
clearly. In this case we lack significant information on the luminosity of the
individual sources, but we gain some more data on the optical depth of the circumstellar
envelopes. We show here the population of the stars in the sample used in the present
investigation on the ([F770W]-[F2550W], $K_{S}$-[F770W]), 
($K_{S}$-[F770W], [F1500W]-[F1800W]), ([F1000W]-[12.5], [F1500W]-[F1800W]), 
([F770W]-[F1000W], [F1500W]-[F2550W]) planes. We base the discussion on the results 
commented in the previous section.

As shown in the top, left panel of Fig.\ref{fccd}, in the 
([F770W]-[F2550W], $K_{S}$-[F770W]) plane, 
low-mass stars (magenta triangles; Sect.~\ref{lowmass}), and massive AGBs
(green squares; Sect.~\ref{hbbstars}), follow similar 
obscuration patterns, the latter extending to redder colours, 
because they reach higher optical depths. The two locii overlap in this case, because 
the main factor affecting the position of the stars is the optical depth; this allows us 
to draw a unique obscuration line. The trend is approximately horizontal up to 
$\tau_{10} \sim 0.1$, [F770W]-[F2550W] $\sim 2$; this behaviour is consistent with the 
arguments presented in Sect.~\ref{lowmass}, related to the modification of the shape 
of the SED, when the main silicate feature appears. For higher $\tau_{10}$, as shown
in the right panel of Fig.~\ref{fspet}, the SED is shifted to the mid-IR, which diminishes 
the flux in the near-IR, thus triggering a significant reddening of the $K_{S}$-[F770W] 
colours: the sequences turn upwards in the diagram. 

The stars with iron dust, both AGB and post-AGB, and those with dual-dust chemistry, 
discussed respectively in Sect.~\ref{festars} and \ref{dualstars}, are located in a diagonal 
sequence above the two afore mentioned obscuration sequences. This is once more due to the particular morphology of their SED, which does not exhibit a clear minimum in the $8~\mu$m region. 


The link between the position of the individual sources and the degree of obscuration
of the circumstellar envelope is clear also in the ($K_{S}$-[F770W], [F1500W]-[F1800W]) plane, shown in
the top, right panel of Fig.\ref{fccd}. Low-mass and massive AGB stars describe obscuration 
sequences that extend to the red, with $K_{S}$-[F770W] and [F1500W]-[F1800W] spanning a range
of $\sim 6$ mag and $\sim 1$ mag, respectively. Among the colours considered so far,
[F1500W]-[F1800W] is the least sensitive to $\tau_{10}$, because, as shown in
Fig.~\ref{fspet}, the presence of the secondary silicate feature at $18~\mu$m is 
accompanied by a general increase in the stellar flux in all the spectral region between 
$10$ and $25~\mu$m: both the [F1500W] and [F1800W] fluxes increase when a prominent 
secondary silicate feature is present in the spectrum. This is the reason why [F1500W]-[F1800W]
never exceeds the maximum value, [F1500W]-[F1800W]$ \sim 0.8$, reached for 
$K_{S}$-[F770W] $\sim 2$.

The peculiar shape of the SED in the $8~\mu$m region makes iron dust and dual dust chemistry stars
 populate regions of the plane below the main obscuration pattern. In particular, AGB stars surrounded
by iron dust are clearly separated from post-AGB stars and low-luminosity objects, with $K_{S}$-[F770W] $\sim 1$, [F1500W]-[F1800W]$ \sim 0.3$. This is due to the steep rise of 
the SED at $\lambda \leq 8 \mu m$ of the iron dust AGB sources respect to their post-AGB counterparts and stars with dual-dust chemistry.
An additional colour where iron post-AGB stars are clearly isolated is the [F1280W]-[F2550W]. This is due to the absence of the silicates secondary feature. This colour in combination with $K_{S}$-[F770W], separates this group of sources from the obscuration sequence traced by the rest of the sample.

The bottom panels of Fig.\ref{fccd} show the distribution of the stars in the sample
considered here in two additional planes proposed by \citet{jones17} to discriminate 
AGB stars from other sources, e.g. young stellar objects and planetary nebulae. 
([F1000W]-[F1280W],[F1500W]-[F1800W]) is also useful to separate O-rich from C-rich AGB 
stars, while ([F770W]-[F1000W],[F1500W]-[F2550W]) is effective in distinguishing between 
galactic and non-galactic sources. Nevertheless, unlike the upper panels diagrams, 
the trend of the colours of the O-rich stars with the optical depth is not straightforward.
This is because the [F1000W] and [F1280W] flux depend on the particular way with which the shape
of the main silicate feature changes with $\tau_{10}$. 

In the ([F1000W]-[F1280W],[F1500W]-[F1800W]) plane the position of the individual stars does
not allow a clear identification of the obscuration properties. The stars with iron
dust and their counterparts with dual dust chemistry can be distinguished more clearly:
indeed their [F1500W]-[F1800W] colours are bluer for a given [F1000W]-[F1280W], because the
depression of the flux at $15~\mu$m is lower in comparison to the stars whose main
dust component is made up of silicates.

In the ([F770W]-[F1000W],[F1500W]-[F2550W]) plane the different classes of stars presented in the
previous section partly overlap and the separation among the various groups is not
straightforward.

\section{Summary and conclusions}
A summary of the overall interpretation of the LMC sources published by \citet{jones12} is 
reported in Table 2. For each source the luminosity, coordinates, periods (when available), 
optical depth at $10~\mu$m, dust mineralogy and evolutionary characterization are given. 
In the case of stars with dual-dust chemistry, we also report the optical depth and the 
percentages of solid carbon and SiC of the inner layer.

The table also includes the interpretation by GS18. We notice an overall satisfactory 
agreement with the latter exploration, but a few sources, for which
the luminosities derived in the present study and in GS18 are substantially
different. For what attains the sources SSID 121 and SSID 4007, the explanation is in the
choice of GS18 to scale the IRS fluxes to match the IRAC and MIPS data. If we
take this into account, the results are consistent. In the other few cases where the results
are not consistent (i.e. SSID 38, 165, 4038 and 4483), we notice that the best fit obtained by
GS18 are indeed of poor quality, as witnessed by the high $\chi^2$ obtained, 
above 750.

The results presented and discussed in the previous sections show that the LMC sample of
the sources classified as evolved M-type AGB stars is mainly composed by two groups: 
a) low-mass stars, evolving through advanced AGB phases; b) massive AGB stars
experiencing HBB. Additional stars included in the sample are low-metallicity massive AGBs
surrounded by mostly iron dust, post-AGB stars and objects with dual-dust chemistry 
composition.
  
Among the two classes of objects listed in points a) and b), massive AGB stars are on 
the average more obscured than low-mass objects; however, the optical
depths of the two groups partly overlap.

Whatever is the combination of {\itshape JWST} fiters adopted to build the observational 
planes, we expect that the separation between the two groups listed in points a) and b) 
can be more appreciated in the colour-magnitude diagrams compared to the colour-colour ones. 
Therefore, on general grounds, the former are a better diagnostic for the characterization 
of the individual sources.

The majority of these stars, with the exception of a few cases discussed in Sect.~
\ref{fepagb}, are characterized by the presence of two silicate features, affecting the
spectral range between $\sim 8$ and $\sim 22~\mu$m. The shape of the SED in this wavelength 
interval depends on the optical depth, in turn related to the presence of dust in the 
circumstellar envelope of the star. As discussed in Sect.~ \ref{spettro}, the modality with 
which the SED changes with $\tau_{10}$ is not 
straightforward. This behaviour makes the filters falling within the afore mentioned
spectral range to be of little help in the interpretation of the observations, because the
obscuration patterns traced in the observational planes built with these filters present
turning points and deviations, which prevents a straight understanding of the results
(see Figs.~\ref{fcmd} and \ref{fccd}).

As discussed in previous sections, a key-quantity in the
interpretations of dusty M-stars is the [F770W] flux. This is because, as far as
the dust around the stars is dominated by the silicate component, the [F770W] flux is
less sensitive to the details of the changes in the shape of the SED with the
degree of obscuration, compared to the filters in the $8~\mu$m $< \lambda < 22~\mu$m
spectral region. On the other hand, the [F770W] flux is extremely sensitive to the
presence of iron dust or, more generally, of featureless dust species and can therefore
be efficiently used to check whether the sample encompasses iron dust stars and 
dual-chemistry sources.

Still within the mid-IR domain, the [F2550W] flux proves an important tool for the 
characterization of the individual sources, as it is sensitive to the overall degree of 
obscuration in the circumstellar envelope. 

Turning to the near-IR, the $K_{S}$ flux is also extremely sensitive to the optical depth,
although the measurements of the flux in the near-IR spectral regions can be hard
for the most extremely obscured sources.

Based on the arguments given above, we conclude that the color-magnitude 
([F770W]-[F2550W], [F770W]) plane is the best diagnostic for dusty M-stars, allowing the 
identification of the following: a) two separated obscuration patterns, holding for massive 
stars undergoing HBB and for their lower mass counterparts; b) stars surrounded by iron dust, 
interpreted as the progeny of low-metallicity massive AGBs; c) post-AGB stars with a 
dual-chemistry dust composition. 

The same two groups of obscured, M-stars can be also distinguished in the color-magnitude 
($K_{S}$-[F770W], [F770W]) plane. In this diagram the post-AGB stars are more clearly 
separated and the $K_{S}$-[F770W] colour of massive AGBs is more sensitive than 
[F770W]-[F2550W] to the degree of obscuration. However, the obscuration pattern traced by 
low-mass dusty AGBs is extremely short and, as discussed previously, the estimate of the 
$K_{S}$ flux is hampered by the presence of large quantities of dust. 

The latter set of filters can also be used to plot the stars in the colour-colour
([F770W]-[F2550W], $K_{S}$-[F770W]) plane. In this case a clear degree of obscuration is
present, despite the sequences of massive AGBs and low-mass stars are not separated. 
Post-AGB stars, iron-dust sources and dual-dust chemistry stars can be distinguished, 
although they populate zones of the plane close to the main obscuration pattern.

Among the colour-colour planes, an alternative possibility to select the different
groups of stars is ($[K_{S}$-[F770W], [F1500W]-[F1800W]).

In summary, future {\itshape JWST} investigations aimed at the characterization of
evolved M stars in galaxies must consider as a priority the collection of data in the 
F770W and F2550W filters. The F210M filter will be also useful to build combinations
of near-IR and mid-IR photometry to distinguish the different obscuration sequences and
characterize the individual sources, although its use is limited by the extremely low
near-IR fluxes of the most obscured stars.

\begin{table*}
\caption{The summary of the overall interpretation of the LMC sources analyzed in this work, including the interpretation by GS18.
This table is available in its entirety in machine-readable form.}                                       
\begin{tabular}{c c c c c c c c c}    
\hline
 SSID  & RA DEC & Period (d)& L/L$_{\odot}$  &  \%(Sil,Al$_{2}$O$_{3}$,Fe) &  $\tau_{10}$  & Type 	&
L/L$_{\odot}$ (GS18) &   \%(Sil,Al$_{2}$O$_{3}$,Fe) (GS18)  \\
\hline
  1	&	69.338  -70.579		&	-	&	7000		&	 (60,40,0)	&	$4.5\times	 10^{-3}$ 	&	Low-mass M	&	5894	&	(40,60,30)  \\
    6	&	72.393  -69.097		&	-	&	18000	&	(15,0,85)	&	$2\times 10^{-3}$ 	&	Iron-dust PAGB		&	-	&	-  \\
 8	& 72.869   -69.930	&	884	&	2100		&	(60,10,30)		&	$3.2\times 10^{-2}$	&	Dual-Dust AGB	&	1660		&	(70,30,30)\\
   	&		&		&	&	C 95, SiC 5	&	$1.5\times 10^{-2}$ 	&	&	&  \\ 
  13	& 73.290   -68.286	& 	916	&	5500		&	-	&	- 	&	No dust	&	5545	  	&	(40,60,1)  \\
 22	&	74.097   -69.463	&	-	&	4700		&	(60,5,35)		&	$1.5\times 10^{-2}$	&	Dual-dust AGB	&	7524		&	(70,30,30)\\
   	&	&	&	&	C 95, SiC 5	&	$8\times 10^{-3}$ 	&	&	&  \\ 
  38	&	76.120   -67.690	&	577	&	34000		&	(65,5,30)	&	$6.7\times	 10^{-2}$ 	&	HBB		&	14409	&	(100,0,30)  \\
  54	&	76.997   -68.657	&	-	&	4300		&	(55,30,15)	&	$1.8\times 10^{-2}$ 	&	Dual-dust AGB		&	3257		&	(70,30,30)  \\  	
     	&	& &	&	C 100	&	$6\times 10^{-3}$ 	&	&	&  \\
\hline       
\hline     
\end{tabular}
\end{table*}

\section*{Acknowledgements}
FDA and DAGH acknowledge support from the State Research Agency (AEI)  
of the  Spanish Ministry of Science, Innovation and Universities  
(MCIU) and the European Regional Development Fund (FEDER) under grant  
AYA2017-88254-P.\\
DK acknowledges the support of the Australian Research Council (ARC) Discovery 
Early Career Research Award (DECRA) grant (95213534).\\
MT acknowledges support the European Research Council (ERC) under the European Union's Horizon 2020 research innovation programme (Grant Agreement ERC-StG 2016, No 716082 'GALFOR', PI: Milone) and from MIUR through the the FARE project R164RM93XW 'SEMPLICE' (PI: Milone).






\bsp	
\end{document}